\documentclass[11pt]{article}

\usepackage[utf8]{inputenc}
\usepackage[T1]{fontenc}
\usepackage{amsfonts,eucal,epsfig,graphicx,float,mathabx}
\usepackage{authblk}
\DeclareMathSymbol{\sm}{\mathbin}{AMSa}{"39}
\DeclareMathSymbol{\Gamma}{\mathalpha}{operators}{"00}
\DeclareMathSymbol{\Pi}{\mathalpha}{operators}{"05}
\usepackage{bbm,bm,stmaryrd,changepage}
\RequirePackage{flushend}
\RequirePackage[numbers,sort&compress]{natbib}
\usepackage[english]{babel}
\setcitestyle{super,open={},close={},comma}
\usepackage{pstricks}

\newcommand{\coinhd}{\raisebox{1pt}{$\Rsh$}}
\newcommand{\coinbd}{$\drsh$}
\newcommand{\coinhg}{\raisebox{1pt}{$\Lsh$}}
\newcommand{\coinbg}{$\dlsh$}
\newcommand{\coindb}{\raisebox{7pt}{\rotatebox{-90}{$\Rsh$}}}
\newcommand{\coindh}{\raisebox{8pt}{\rotatebox{-90}{$\Lsh$}}}
\newcommand{\coingh}{\raisebox{8pt}{\rotatebox{-90}{$\dlsh$}}}
\newcommand{\coingb}{\raisebox{7pt}{\rotatebox{-90}{$\drsh$}}}
\newcommand{\rtd}{\sqrt{\phantom{\!\bar{\hat\gamma}}}\hspace{-5pt}\frac32}
\newcommand{\ii}{\mathbbm{i}}
\newcommand{\e}{\mathop{\,\rm e}\nolimits}
\newcommand{\At}{\mathop{@}\limits}
\newcommand{\mo}{m_{\rm o}}
\newcommand{\Id}{{\,\rm Id}}

\newcommand{\sign}{{\rm sign}}
\newcommand{\Sym}{{\rm Sym}}
\newcommand{\mod}{{\rm mod}\;}
\newcommand{\R}{\mathbb{R}}

\newcommand{\Z}{\mathbb{Z}}

\newcommand{\gt}{{\mathop{\widetilde\gamma}}}
\title{Topology of contact points in Lieb-kagomé model}
\author{G. Abramovici}
\affil{Université Paris-Saclay, CNRS, Laboratoire de Physique des Solides,
91405, Orsay, France\\
abramovici@lps.u-psud.fr}

\begin{document}
\maketitle
\abstract{
We analyse Lieb-kagomé model, a three-band model with contact points showing
particular examples of the merging of Dirac contact points. We prove that
eigenstates can be parametrized in a \textit{classification} surface, which is a
hypersurface of a 4-dimension space. This classification surface is a powerful
device giving topological properties of the energy band structure; the analysis
of its fundamental group proves that all singularities of the band structure can
be characterized by four independent winding (integer) numbers.  Lieb case
separates: its \textit{classification} surface differs and there is only one
winding number.}

%\keywords{Singularities \and protected states \and homotopy classification}

\section{Introduction}

In all times, physical systems have been investigated through geometry. In
recent literature, such a mathematical approach has become essential:
quantification numbers are protected by \textbf{topological} properties, like
the quantum of magnetic flux associated to the quantum anomalous Hall
effect,\cite{Haldane} or particular quantum states by \textbf{singularities},
like zero-mass Dirac states in graphene\cite{Bena,FuchsEpjB,Lih2015} or
zero-energy Majorana states in superconducting
systems.\cite{Read,Kitaev,FuKaneMele} At first, topological classification has
been performed in real space,\cite{Dirac,AharonovBohm} then in reciprocal
space.\cite{Berry,Volovik} The mother of all such characterization would be to
map physical states on a universal surface, called \textit{classification}
surface, capturing all topological properties. This has been done indeed for
two-band systems, where the standard classification surface\cite{Bloore} is
sphere $S_2$, called Poincaré space\cite{Poincare} or Bloch
sphere.\cite{Arecchi,Lieb} 

We provide here the classification surface for a three-band model. We study
Lieb-kagomé model, which addresses the merging of contact points between energy
bands.\cite{Montambaux,Lih2012} We focus in particular on Dirac contact points,
which have been observed in several physical systems.\cite{Apaja,Nathan,Xiao}
Here, contact points are the singularities of the energy band structure in
reciprocal space.  In order to characterize these singularities, one must
determine the surface in which all free parameters, which determine the
corresponding eigenstates, can be embedded.\cite{VolovB} This surface is
indeed the \textit{classification} surface.

Lieb-kagomé model\cite{Tsai,Lih-King} interpolates Lieb and kagomé ones,
with interpolating parameter $t'\in[0,1]$.  For $t'=0$, it is equal to Lieb
model; for $t'=1$, it is equal to kagomé one. One of our aim and interest in
this model is to understand the topological classification of both Lieb and
kagomé models by setting parameter $t'$ in the vicinity of either Lieb limit,
$t'\sim0$, or kagomé one, $t'\sim1$.

The band structure of Lieb model has been already studied\cite{Lieb} and reveals
a unique contact point between three bands simultaneously: the energy spectrum
at this point has a triple degeneracy, the upper and lower bands show typical
straight cones, while the middle one is flat at the intersection point.  It has
already been established\cite{Goldman,Tsai} that the classification surface of
Lieb model is $\mathcal{S}_1$, the ordinary circle embedded in the plan; thus,
the topology of the energy band singularities in Lieb model is classified by the
first homotopy group $\pi_1({\cal S}_1)=\Z$. This determination is however
local, it was not possible, using previous method that were based on an improved
two-band analysis,\cite{Lih-King} to predict the relation between the winding
around a singularity and that around another one.  In other words, it was not
possible to find the periodicity of the winding number in reciprocal space,
except of what concerns close Dirac points about to merge at $t'=0$.  Using a
three-band resolution, we will rebuild this winding number and determine its
complete periodicity. 

In kagomé model, using again a two-band approach, it is only possible to deduce,
from the study of contact point aggregates, partial results concerning the
periodicity between close Dirac points about to merge.\cite{Lih-King} Moreover,
it is not possible, with such an approach, to determinate its classification
surface.  Instead, two-band approach only reveals ${\cal S}_1$-like
surfaces.\cite{kagome,Chen} With the determination of the exact classification
surface, we will show that four winding numbers can be defined, that give
different scenarios for the aggregation of Dirac points.

In summary, we have determined the classification surface for all
$t'{\in}[0{,}1]$. Case $t'=0$ separates. For $t'>0$, one finds a unique
hypersurface $\cal S$ embedded in a 4-dimension space, which we call
\textit{universal} classification surface.  We have exhibited a tridimensional
representation of $\cal S$, with 18 toroidal holes. Its fundamental group is
complicated and verifies $\pi_1({\cal S})\ge \Z_{12}$. However, one needs not
its determination, since eigenstate parameters do not spread over the whole
surface $\cal S$ but only over a part of it, which we call \textit{effective}
classification surface with 12 holes inside.  Eventually, we will prove that
\textbf{four} winding numbers, called $\omega_2$, $\omega_3$, $\omega_4$ and
$\omega_5$, are sufficient to describe the topological properties of the energy
band structure. We also introduce $\omega_1=\frac{\omega_4}4$, which deals with
Lieb limit.  Each $\omega_i$ has specific periodicity (in reciprocal space) and
properties.  For instance, $\omega_2$ and $\omega_4$ are periodic with
$(2\pi,2\pi)$, $(4\pi,0)$ and $(0,4\pi)$ translations, while $\omega_3$ is
periodic with $(4\pi,0)$ and $(0,2\pi)$ translations and $\omega_5$ is periodic
with $(2\pi,0)$ and $(0,2\pi)$ ones.

Before entering into complicated mathematical considerations, the first and
simple task to do is to enumerate exactly all degrees of freedom in the
system.\cite{Asano} In the general case, eigenvectors can be described by six
real degrees of freedom: this can be understood as a result of the general
Jordan decomposition applied to a hermitian hamiltonian, which is described by
nine real parameters, from which the three real eigenvalues must be subtracted;
it can be either understood by hand: each eigenvector has three complex
coordinates, i.e. six real parameters, but one must subtract two degrees of
freedom (the normalization and overall-phase); moreover, from the remaining
$3\times4$ real parameters of the three eigenvectors, six must be discarded,
that express orthogonality relations between eigenvectors. 

Time-inversion symmetry gives $\overline{H({-}{\bf k})}=H({\bf k})$ while
inversion symmetry gives $H({-}{\bf k})=H({\bf k})$, which altogether discards
in our case three degrees of liberty; this can also be understood by hand: all
eigenvector coordinates prove real, so each eigenvector has three real
parameters, from which one degree of freedom must be reduced (the
normalization); eventually, from the remaining $3\times2$ real parameters of the
three eigenvectors, three must be discarded, that express orthogonality
relations. We are left with \textbf{3 degrees of freedom}, which fits hunter's
rule\cite{ToulouseKleman} and is their actual number.

In this article, we will first describe Lieb-kagomé model in detail, in
particular we give the representation of eigenstates in terms of projectors,
secondly we will determine universal classification surface $\cal S$, then we
will analyse the relevant part of its fundamental group, introducing
classification surfaces $\widetilde{\cal S}_1$ and  ${\cal E}={\cal S}_2\times%
{\cal R}\times{\cal C}\times\cal T$. We will examine Lieb and kagomé cases
separately.  Afterwards, we will present our complete results in terms of
winding numbers.  Eventually, we discuss them and conclude.

\subsection{Notations}

The reader must be careful not to make confusions between parameters $r$, $t$ or
$s$ (the latter to be introduced in appendix only) and projectors $r$, $t$ or
$s$. Symbol $\gamma$ is exclusively for paths in reciprocal space, $\gt$ for
their images in $\cal S$ and all paths in other surfaces (resp.  ${\cal Q}_a$,
${\cal Q}_d$, ${\cal S}_2$, ${\cal R}'$, $\cal C$ and $\cal T$) are denoted with
the corresponding projection (resp. $\tau_a$, $\tau_d$, $s$, $r$, $c$ and $t$).
Number $i$ in $a^i$ and $b^i$ is always an index, while power $n$ are written as
$(a^i)^n$.

We have used a color code: green for diagonal contact points and red for
antidiagonal ones, which we have extended as much as possible through the whole
article.

\section{Lieb-kagomé model}

\paragraph{Atomic structure} In Lieb-kagomé model, there are three types of
atoms, with equal stoichiometry. Next-near neighbours are interacting with equal
intensity 1.  Second-near neighbours of type two and three interact one another
with intensity $0\le t'\le1$, according to the scheme in Fig.~\ref{atomic}:
\begin{figure}[H]
\begin{center}
\begin{pspicture}(-0.3,-0.3)(2.3,2.3)
\pscircle[fillstyle=solid,fillcolor=blue,linecolor=blue](0,0){0.1}
\pscircle[fillstyle=solid,fillcolor=blue,linecolor=blue](0,2){0.1}
\pscircle[fillstyle=solid,fillcolor=blue,linecolor=blue](2,0){0.1}
\pscircle[fillstyle=solid,fillcolor=blue,linecolor=blue](2,2){0.1}
\pscircle[fillstyle=solid,fillcolor=yellow,linecolor=black](0,1){0.1}
\pscircle[fillstyle=solid,fillcolor=yellow,linecolor=black](2,1){0.1}
\pscircle[fillstyle=solid,fillcolor=red,linecolor=red](1,0){0.1}
\pscircle[fillstyle=solid,fillcolor=red,linecolor=red](1,2){0.1}
\psline(0,0)(0,2)
\psline(0,0)(2,0)
\psline(0,2)(2,2)
\psline(2,0)(2,2)
\psline[linestyle=dashed](1,0)(2,1)
\psline[linestyle=dashed](0,1)(1,2)
\end{pspicture}
\end{center}
\caption{Atomic structure of Lieb-kagomé model. First-near links are represented
by solid lines, second-near ones by dashed lines.}
\label{atomic}
\end{figure}
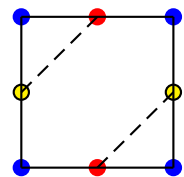

\paragraph{Bloch hamiltonian} The corresponding Bloch hamiltonian (here given in
basis II, see the representation in basis I in appendix\cite{Bena}) is
\[
H=\pmatrix{0&2\cos\frac{k_x}2&2t'\cos\frac{k_x+k_y}2\cr
2\cos\frac{k_x}2&0&2\cos\frac{k_y}2\cr
2t'\cos\frac{k_x+k_y}2&2\cos\frac{k_y}2&0\cr}.
\]
For $t'=0$, $H$ is equal to Lieb Bloch hamiltonian. For $t'=1$, $H$
corresponds to kagomé model with a deformed crystalline structure. 

%$H$ is real. 
This model is time-reversal invariant: $\overline{H(-\mathbf{k})}
=H(\mathbf{k})$ and inversion-invariant: $(-{\cal I}) H(-\mathbf{k})(-{\cal I})
=H(\mathbf{k})$, where $\cal I$ is the identity matrix.  Both symmetries follow
the reality of $H$ and its even parity as a function of $\mathbf{k}$. $H$ has a
double periodicity: $H(k_x+4\pi,k_y)=H(k_x,k_y)$ and
$H(k_x,k_y+4\pi)=H(k_x,k_y)$. The corresponding effective Brillouin zone is
four time larger than the real one.

\paragraph{Energy spectrum} We use a specific notation for the three
corresponding energies $e_{\sm1}$ (lower band), $e_1$ (middle band) and $e_0$
(upper band). They write
\[
e_n(k_x,k_y)=2(1-2n^2)\sqrt{\frac{2\,p(k_x,k_y)}3}
\cos\bigg(\frac{\Theta(k_x,k_y)+n\pi}3\bigg),
\]
where
\[
\Theta(k_x,k_y)=\left\{
\matrix{\tan^{-1}\big[r(k_x,k_y)\big]^{\phantom{k^k}}&
\hbox{when }q(k_x,k_y)>0\;, \cr
\pi-\tan^{-1}\big[r(k_x,k_y)\big]&\hbox{when }q(k_x,k_y)<0\;,\cr
\displaystyle\frac{\phantom{c}}2\!\!\!\!{^{\displaystyle\pi}}&
\hbox{when }q(k_x,k_y)=0\;,\cr}
\right.
\]
with
\begin{eqnarray*} 
p(k_x,k_y)&=&2+t^{\prime 2}+\cos(k_x)+\cos(k_y)+t^{\prime 2}\cos(k_x+k_y)\,,\\
q(k_x,k_y)&=&1+\cos(k_x)+\cos(k_y)+\cos(k_x+k_y)\,,\\
\hbox{and}&&
r(k_x,k_y)=\sqrt{-1+\frac{2p(k_x,k_y)^3}{27t^{\prime 2}q(k_x,k_y)^2}}\;.
\end{eqnarray*}

\begin{figure}[H]
\begin{center}
\includegraphics[width=5cm]{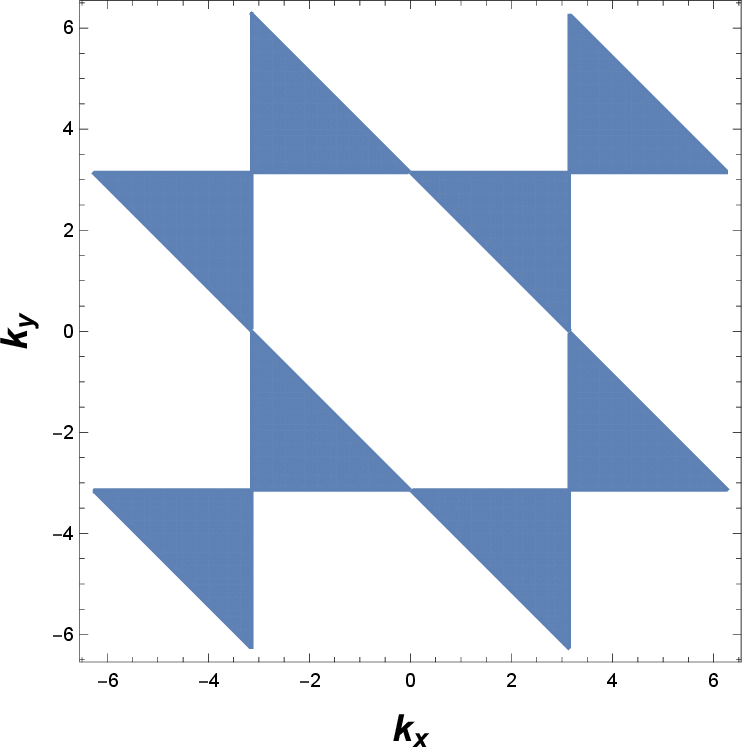}
\end{center}
\caption{Sign of $q(k_x,k_y)=1+\cos(k_x)+\cos(k_y)+\cos(k_x+k_y)$. Dark area
corresponds to ${-}1$.}
\label{signarctan}
\end{figure}
\begin{figure}[H]
\begin{center}
\includegraphics[width=6cm]{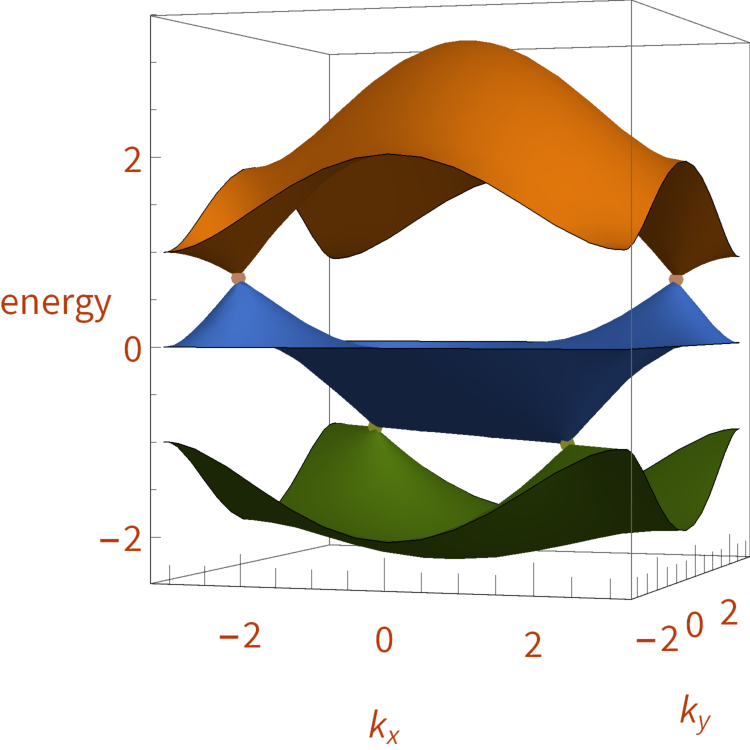}
\end{center}
\caption{Energy bands in the Brillouin zone $]{-}\pi,\pi]\times]{-}\pi,\pi]$ for
$t'=\frac12$. The four contact points are artificially enlarged.}
\label{bandes}
\end{figure}

The sign of $q(k_x,k_y)$ is shown in Fig.~\ref{signarctan}. $\forall(k_x,k_y)$,
$e_{\sm1}\le e_1\le e_0$ is verified by energy bands, represented in
Fig.~\ref{bandes}.
\noindent

\begin{figure}[H]
\begin{center}
\includegraphics[width=8cm]{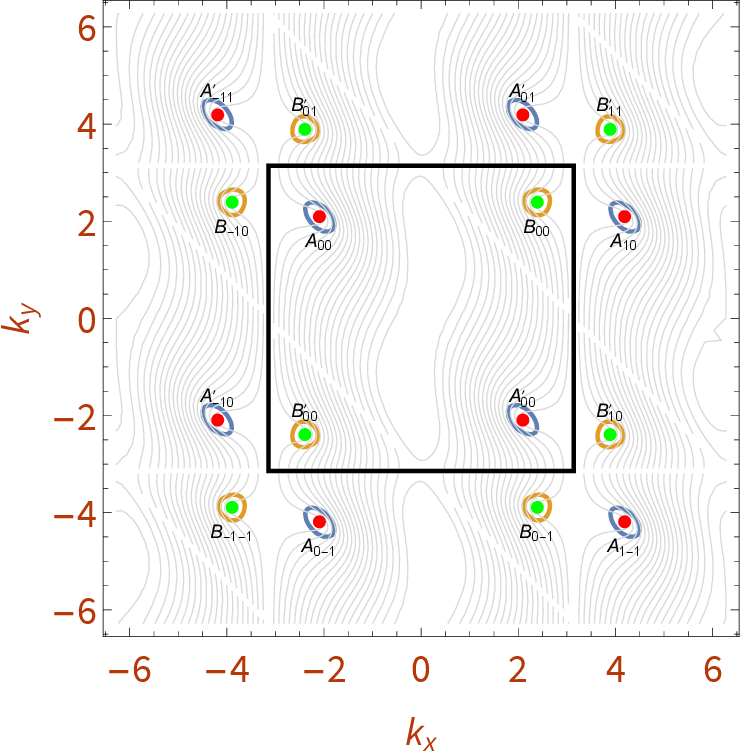}
\end{center}
\caption{Contour lines of $e_{\sm1}-e_1$ and $e_1-e_0$ and positions of all
contact points in the reciprocal space zone $]{-}2\pi,2\pi]\times
]{-}2\pi,2\pi]$ for $t'=\frac12$. In black, Brillouin zone $]{-}\pi,\pi]\times
]{-}\pi,\pi]$ boundary.}
\label{intersect}
\end{figure}
 
For $0<t'<1$, there are exactly four Dirac contact points per Brillouin zone,
which can be grouped into couples: we write $B$, $B'$, Dirac points that are
located in the diagonal $k_x=k_y\ (\mod 2\pi)$ and correspond to contact points
between the two upper bands $e_1$ and $e_0$; we write $A$, $A'$, those located
in the antidiagonal $k_x={-}k_y\ (\mod 2\pi)$ and which correspond to contact
points between the two lower bands $e_{\sm1}$ and $e_1$.

Taking into account $(2\pi,0)$ and $(0,2\pi)$ translations, contact points range
in the whole reciprocal space. Those of $A$, $A'$ kind write $A_{mn}{=}
(\alpha_{t'}+2\,\!m\,\!\pi,{-}\alpha_{t'}+2\,\!n\,\!\pi)\!\!$\penalty-10000 and
$A'_{mn}{=}({-}\alpha_{t'}+2\,\!m\,\!\pi,\alpha_{t'}+2\,\!n\,\!\pi)$, with
$\alpha_{t'}= 2\cos^{-1}(t')$ for any integers $(m,n)\in\Z^2$. Those of $B$,
$B'$ kind write
$B_{mn}{=}(\beta_{t'}{+}2\,\!m\,\!\pi,\beta_{t'}{+}2\,\!n\,\!\pi)$ and
$B'_{mn}{=}({-}\beta_{t'}{+}2\,\!m\,\!\pi,$\penalty-10000
${-}\beta_{t'}{+}2\,\!n\,\!\pi)$, with
$\beta_{t'}=2\cos^{-1}\!\Big(\frac{\sqrt{1+8t^{\prime 2}}-1}{4t'}\Big)$ for any
integers $(m,n)\in\Z^2$. We show some of these points for $t'=\frac12$ in the
following representation Fig.~\ref{intersect} of energy differences in an
extended reciprocal space zone:

Contact points move while $t'$ describes interval $[0,1]$. For $t'=0$, one finds
$\alpha_0=\beta_0=\pi$, so points $A_{00}$, $A'_{\sm11}$, $B_{\sm10}$ and
$B'_{01}$ merge altogether into a unique singularity at point $M_{00}$, where we
define $M_{mn}=((2m+1)\pi,$\penalty-1000$(2n+1)\pi)$. Similar mergings occur modulo $2\pi$ in
each $k_x$ and $k_y$ directions. For $t'=1$, $\beta_1=\frac{2\pi}3$ and
$\alpha_1=0$, so points $A_{00}$ and $A'_{00}$ merge into a unique singularity
at point $\Gamma_{00}$, where we define $\Gamma_{mn}=(2m\pi,2n\pi)$.  Similar
mergings occur modulo $2\pi$ in each $k_x$ and $k_y$ directions. One can also
mention interesting value $\alpha_{\frac12}=\frac{2\pi}3$.  This is resumed in
Fig.~\ref{fsing}:
\begin{figure}[H]
\begin{center}
\includegraphics[width=4cm]{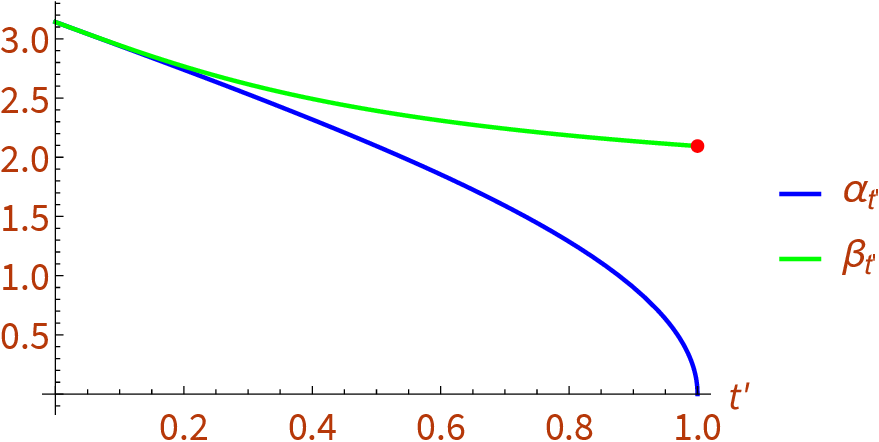}
\end{center}
\caption{Curves of $\alpha_{t'}$ and $\beta_{t'}$ versus $t'$.}
\label{fsing}
\end{figure}

\paragraph{Eigenstates} We write $|v_n\rangle$ each eigenvector corresponding to
energy $e_n$, $\forall n={-}1,0,1$. The column vector writes
\[
\!\!\!\!\!\!\!\!\!\!
v_n(k_x,k_y)=\pmatrix{\cos\!\big(\!\frac{k_x}2\!\big)
\cos\!\big(\!\frac{k_y}2\!\big)
+\frac{t'}2\cos\!\big(\!\frac{k_x+k_y}2\!\big)e_n(k_x,k_y)\cr
{-}\cos\!\big(\!\frac{k_x}2\!\big)^2+\frac14e_n(k_x,k_y)^2\cr
t'\cos\!\big(\!\frac{k_x}2\!\big)\cos\!\big(\!\frac{k_x+k_y}2\!\big)
+\frac12\cos\!\big(\!\frac{k_y}2\!\big)e_n(k_x,k_y)\cr}.
\]

These expressions are not normalized, so one needs to use $|v_n\rangle/
\sqrt{\langle v_n|v_n\rangle}$ in order to get normalized vectors.  To each
$|v_n\rangle$ corresponds a projector $\Pi_n= \frac{|v_n\rangle\langle
v_n|}{\langle v_n|v_n\rangle}$. We decompose $\Pi_n$ into Bloch components
$a^1_n$, ..., $a^8_n$ such that:
\[
\Pi_n=\frac13{\cal I}+\frac1{\sqrt3}\sum_{i=1}^8 a^i_n\lambda_i
\]
with the Gell-Mann matrices:
\newline
\begin{tabular}{cc}\\
$\lambda_1=\pmatrix{0&1&0\cr1&0&0\cr0&0&0\cr}$;& 
$\lambda_2=\pmatrix{0&{-}\ii&0\cr\ii&0&0\cr0&0&0\cr}$;\\ 
$\lambda_3=\pmatrix{1&0&0\cr0&{-}1&0\cr0&0&0\cr}$;& 
$\lambda_4=\pmatrix{0&0&1\cr0&0&0\cr1&0&0\cr}$; \\
$\lambda_5=\pmatrix{0&0&{-}\ii\cr0&0&0\cr\ii&0&0\cr}$;& 
$\lambda_6=\pmatrix{0&0&0\cr0&0&1\cr0&1&0\cr}$;\\
$\lambda_7=\pmatrix{0&0&0\cr0&0&{-}\ii\cr0&\ii&0\cr}$;& 
$\lambda_8=\displaystyle\frac1{\sqrt3}\pmatrix{1&0&0\cr0&1&0\cr0&0&{-}2\cr}$.
\end{tabular}

\section{Determination of the universal classification surface}

Calculations with projectors provide phase independent results, contrary to
vector based ones. We will, in particular, ensure that they account for the
correct number of degrees of freedom, altogether.

\subsection{One projector conditions}

Let us introduce equation
\begin{equation}
\label{rel1}
a.a=1\quad\hbox{and}\quad a\star a=a\;,
\end{equation}
where $a=(a^1,...,a^8)$ is the Bloch octuplet describing an eigenstate,
$a.b=\sum_{i=1}^8a^i b^i$ is a scalar product and the star product $a\star b$ is
defined by $a\star b=\sum_{i=1}^8 a^j b^k d_{ijk}$ with
\begin{eqnarray*}
&&d_{ii8}=d_{i8i}=d_{8ii}=\frac1{\sqrt3}\qquad \forall i=1,2,3\;;\\ 
&&d_{ii8}=d_{i8i}=d_{8ii}={-}\frac1{2\sqrt3}\quad \forall i=4,..,7\;;\\ 
&&d_{888}={-}\frac1{\sqrt3}\;;\\
&&d_{146}=d_{461}=d_{614}=d_{164}=d_{641}=d_{416}=\frac12\;;\\
&&d_{157}=d_{571}=d_{715}=d_{175}=d_{751}=d_{517}=\frac12\;;\\
&&d_{256}=d_{562}=d_{625}=d_{265}=d_{652}=d_{526}=\frac12\;;\\
&&d_{344}=d_{434}=d_{443}=d_{355}=d_{535}=d_{553}=\frac12\;;\\
&&d_{247}=d_{472}=d_{724}=d_{274}=d_{742}=d_{427}={-}\frac12\;;\\
&&d_{366}=d_{636}=d_{663}=d_{377}=d_{737}=d_{773}={-}\frac12\;.
\end{eqnarray*}
Each projector $\Pi_n$ respects\cite{Goyal} (\ref{rel1}), with $a=a_n$,
$\forall n={-}1,0,1$.

Let us study (\ref{rel1}) formally. After some tedious calculations, one finds
that it is equivalent to the five equations
\begin{eqnarray}
\nonumber
(a^4)^2+(a^5)^2+(a^6)^2+(a^7)^2&&\\
\label{eq1}
+\frac43\Big(a^8+\frac14\Big)^2&=&\frac34\;;\\
\label{eq1b}
(a^1)^2+(a^2)^2+(a^3)^2&=&\frac13(a^8+1)^2\;;\\
\label{eq2}
\sqrt3(a^4a^6+a^5a^7)&=&(1-2a^8)a^1\;;
\\
\label{eq2b}
\sqrt3(a^5a^6-a^4a^7)&=&(1-2a^8)a^2\;;
\\
\sqrt3\big((a^4)^2+(a^5)^2-(a^6)^2-(a^7)^2\big)&=&2(1-2a^8)a^3.
\label{eq3}
\end{eqnarray}
All coordinates in $v_n(k_x,k_y)$ are actually real, so there is no imaginary
term in their Bloch decomposition. Therefore, $a^2=0$, $a^5=0$ and $a^7=0$ for
all $a=a_n$, with $n=-1,0,1$, \textbf{and we discard these three components in
the whole article hereafter}. Equations (\ref{eq1}), (\ref{eq2}) and (\ref{eq3})
become
\begin{eqnarray}
\label{eq1a}
(a^4)^2+(a^6)^2+\frac43\Big(a^8+\frac14\Big)^2&=&\frac34\;;\\
\label{eq2a}\sqrt3a^4a^6&=&(1-2a^8)a^1\;;\\
\sqrt3\big((a^4)^2-(a^6)^2\big)&=&2(1-2a^8)a^3\;.
\label{eq3a}
\end{eqnarray}

We have skipped (\ref{eq1b}), which becomes redundant and (\ref{eq2b}), which
becomes trivial. Altogether, for each projector $\Pi_n$, there are five degrees
of freedom, constrained to the three equations (\ref{eq1a}), (\ref{eq2a}) and
(\ref{eq3a}) applied with $a=a_n$.  Therefore, we are left with two degrees for
each projector. In the following, we will keep $a^4_n$ and $a^6_n$ for
each $n={-1},0,1$, using (\ref{eq1a}) to express $a^8_n$ in terms of $a^4_n$ and
$a^6_n$, then (\ref{eq2a}) to express $a^1_n$ in terms of $a^4_n$ and $a^6_n$
and similarly (\ref{eq3a}) to express $a^3_n$ in terms of $a^4_n$ and $a^6_n$.
We are left with $6=3\times2$ components,
$\{a^4_{\sm1},a^6_{\sm1},a^4_0,a^6_0,a^4_1,a^6_1\}$, which exceeds three, the
maximal number of free parameters. One must now take orthogonality relations
into account in order to discard irrelevant ones.

\subsection{Two projector conditions}

For each projector $\Pi_n$, with $n={-}1,0,1$, the standard relations $\Pi_n
\Pi_n=\Pi_n$ are already embedded in relations (\ref{eq1a}), (\ref{eq2a}) and
(\ref{eq3a}). We will now examine mutual relations between projectors.

\subsubsection{Completeness relation}

First of all, they follow
\begin{equation}
\label{fermeture}
\Pi_{\sm1}+\Pi_1+\Pi_0=\Id,
\end{equation}
so one of them is determined by the others; (\ref{fermeture}) writes
\begin{equation}
\label{afermeture}
a^4_p=-a^4_m-a^4_n\quad\hbox{and}\quad a^6_p=-a^6_m-a^6_n\;,
\end{equation}
where $\{m,n,p\}$ is a permutation of $\{{-}1,0,1\}$; in other words
$m=\sigma({-}1)$, $n=\sigma(0)$ and $p=\sigma(1)$ with
$\sigma\in\Sym(\{-1,0,1\})$ an arbitrary permutation. (\ref{afermeture})
discards two free parameters, so the actual counting of degrees of freedom
already drops to four.

\subsubsection{Other relations reduce to a unique equation}

All other relations will reduce to a unique equation, which we introduce at
once: consider any arbitrary octuplets $a$ and $b$, it writes
\begin{equation}
\label{rel2}
a.b={-}\frac12\quad\hbox{and}\quad a\star b={-}a-b\;.
\end{equation}

We will prove now that condition: ``$\forall i\ne j$, $\Pi_i.\Pi_j=0$'' is
equivalent to (\ref{rel2}), with $a=a_m$ and $b=a_n$, where $\{m,n,p\}$ are
defined above and keeping in mind that all components $a^i_p$ can express in
terms of $(a^4_m,a^6_m,a^4_n,a^6_n)$, thanks to (\ref{afermeture}).

Let us first study (\ref{rel2}) formally. This again gives tedious calculations;
several particular cases arise, which separate from the general solution. Let us
give two of them. The first one leads to equations
\begin{equation}
\label{a8b8s2}
\left\{
\begin{tabular}{l}
$a^8=b^8=\frac12$,\quad $a^4=a^6=b^4=b^6=0$,\\
 $b^1={-}a^1$,\quad$b^3={-}a^3$\\
and $(a^1)^2+(a^3)^2=\frac34$.
\end{tabular}
\right.
\end{equation}
The second one leads to equations
\begin{equation}
\label{a8b8s4}
\left\{
\begin{tabular}{l}
$a^8=b^8={-}\frac14$,\quad $b^4={-}a^4$,\quad $b^6={-}a^6$,\\ $b^1=a^1=
\displaystyle\frac{2a^4a^6}{\sqrt3}$,\quad $b^3=a^3=\displaystyle%
\frac{(a^4)^2-(a^6)^2}{\sqrt3}$\\
and\quad$(a^4)^2+(a^6)^2=\frac34$.
\end{tabular}
\right.
\end{equation}
Other cases exist, which study is irrelevant here. Keeping Lieb case $t'=0$
apart, which needs special investigations, all particular situations only arise
for $k_x=k_y\ (\mod\;\pi)$ or for $k_x={-}k_y\ (\mod\;\pi)$, at all $t'>0$.
These lines merge into the plan, letting all components behave analytically, so
we can ignore them. Let us eventually give the generic solution of (\ref{rel2});
we first introduce coordinates $X$, $Y$ and $Z=V+W$:
\begin{eqnarray*}
X(a^4,a^6,b^4,b^6)&=&a^4b^4+a^6b^6\;,\\
Y(a^4,a^6,b^4,b^6)&=&a^6b^4-a^4b^6\;,\\
V(a^4,a^6,b^4,b^6)&=&(a^4)^2+(b^4)^2\\\hbox{and}\quad
W(a^4,a^6,b^4,b^6)&=&(a^6)^2+(b^6)^2\;.
\end{eqnarray*}
Then, the generic solution of (\ref{rel2}) writes
\begin{equation}
3X^2Z+6X^3+3XY^2+Y^4=0\;,
\label{unique}
\end{equation}
while all other components are given by
\begin{eqnarray*}
a^1={-}\frac{a^4a^6Y^2}{\sqrt3X(X+V)}\;,&&
b^1={-}\frac{b^4b^6Y^2}{\sqrt3X(X+W)}\;,\\
a^3=\frac{((a^6)^2-(a^4)^2)Y^2}{2\sqrt3X(X+V)}\;,&&
b^3=\frac{((b^6)^2-(b^4)^2)Y^2}{2\sqrt3X(X+W)}\;,\\
a^8=\frac{(3X+W)V+2X^2}{2Y^2}\;,&&
b^8=\frac{(3X+V)W+2X^2}{2Y^2}\;,
\end{eqnarray*}
where the dependency in $(a^4,a^6,b^4,b^6)$ is hidden.
Taking into account (\ref{unique}) with $a=a_m$ and $b=a_n$, 
we have proven that the number of degrees of freedom is exactly three. 

\section{Classification of singularities}

Following previous work,\cite{VolovB,Tsai,Avron,Lih-King} we characterize
topological singularities of the energy band structure by mapping closed paths
from reciprocal space onto surface $\cal S$. This mapping writes $\gamma\to
\gt$, where $\gamma$ is defined in reciprocal space and $\gt$ in $\cal S$,
and defines a subgroup of the fundamental group $\pi_1({\cal S})$.

\subsection{Description of universal classification surface $\cal S$}

Let $(m,n,p)$ be some permutation of $\{-1,0,1\}$ as already defined, we recall
that (\ref{afermeture}) allows one to skip all $a^i_p$ components, so all Bloch
components express in terms of $(a^4_m,a^6_m,a^4_n,a^6_n)$, while (\ref{rel2}),
applied with $a=a_m$ and $b=a_n$, dismissing non generic cases, proves that
these components are not free and live in surface $\cal S$, defined by
(\ref{unique}), which writes explicitly
\begin{eqnarray*}
3(a^4_ma^4_n+a^6_ma^6_n)^2
\big((a^4_m)^2{+}(a^4_n)^2{+}(a^6_m)^2{+}(a^6_n)^2\big)\\
+6(a^4_ma^4_n+a^6_ma^6_n)^3+3(a^4_ma^4_n+a^6_ma^6_n)\times\\
(a^6_ma^4_n-a^4_ma^6_n)^2+(a^6_ma^4_n-a^4_ma^6_n)^4=0\;.&&
\end{eqnarray*}

$\cal S$ is a tridimensional surface embedded in the 4-dimension space spanned
by components $(a^4_m,a^6_m,a^4_n,a^6_n)$. It does not depend on $t'$. A direct
description seems, at first, not available but we have been lucky enough to find
out that there is a singularity at point $O=(0,0,0,0)$. Its determination is
explained in appendix, however, its properties have been redundantly proven
afterwards, as will now be explained.

\begin{figure}[H]
\begin{center}
\includegraphics[width=4cm]{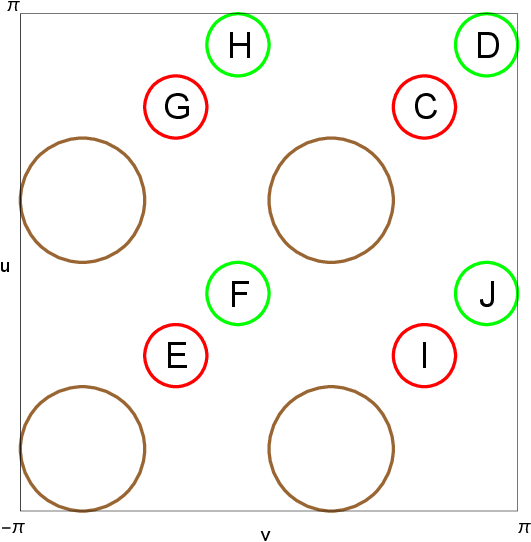}
\end{center}
\caption{Schematic representation of the $t$-axis view of $\cal S$, with the
definition of labels $C$ to $J$.}
\label{stsk}
\end{figure}

Several 4-dimensional connected volumes, cone-shaped, lying \textbf{outside} of
$\cal S$, point towards the origin $O$. They form \textbf{holes} joining at $O$.
To explore this net of holes, we draw the intersection of $\cal S$ with
$r$-$S_3$, the 3-sphere of radius $r$, using Hopf coordinates:
\begin{eqnarray*}
a^4_m&=&r\cos t\cos u\;,\\
a^4_n&=&r\cos t\sin u\;,\\
a^6_m&=&r\sin t\cos v\;,\\
a^6_n&=&r\sin t\sin v\;,
\end{eqnarray*}
with $0\le t<\pi/2$, ${-}\pi<u\le\pi$ and ${-}\pi<v\le\pi$.  This intersection
does not vary with $r$, when $r$ ranges interval $]0,\rtd]$.  One
observes 18 toroidal holes, as seen in Fig.~\ref{St}. Detailed folding rules are
given in appendix, as well as the counting of all holes. 12 holes are parallel
to $t$-axis, 3 parallel to $u$-axis and 3 to $v$-axis. 
\begin{figure}[H]
\begin{center}
\includegraphics[width=4cm]{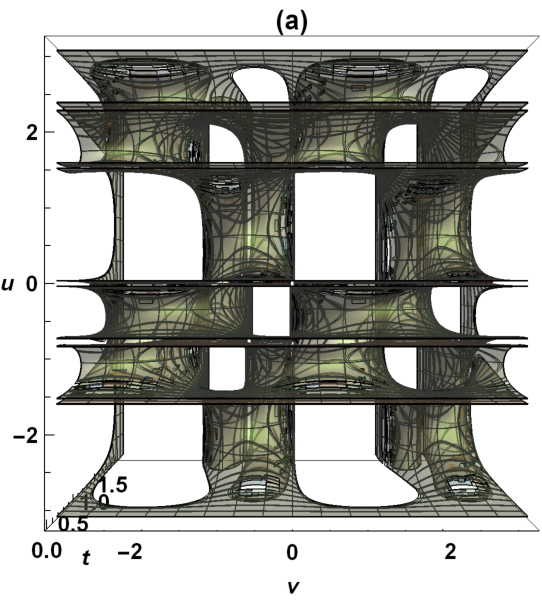}
\includegraphics[width=4cm]{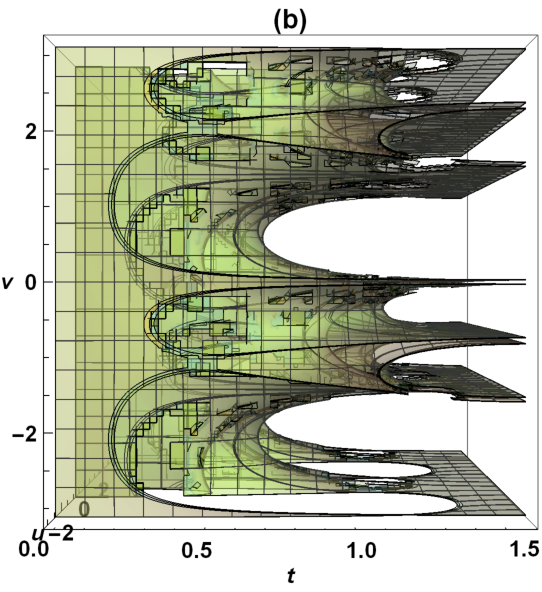}
\end{center}
\caption{Intersection of $\cal S$ with $S_3$. This figure does not change when
the radius of $S_3$ is rescaled from 0 to $\rtd$. In (a), one observes
12 holes from the $t$-direction. In (b) we show a view from the $u$-direction
but a symmetrical image would be obtained from the $v$-direction. Another image
is given in appendix to help for the correct counting of holes.}
\label{St}
\end{figure}
\begin{figure}[H]
\begin{center}
\includegraphics[width=4cm]{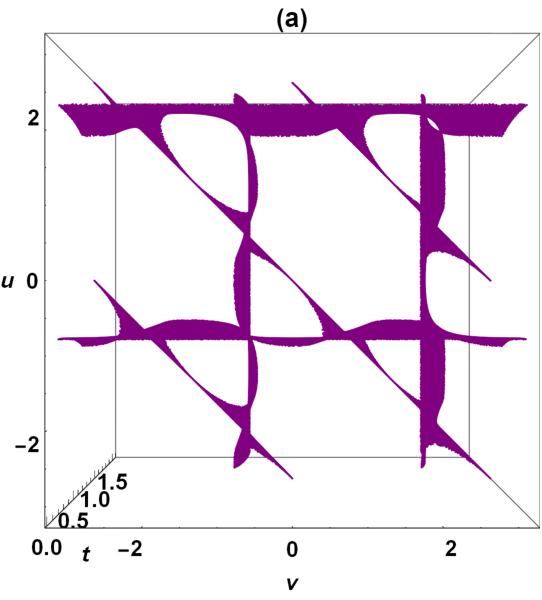}
\includegraphics[width=4cm]{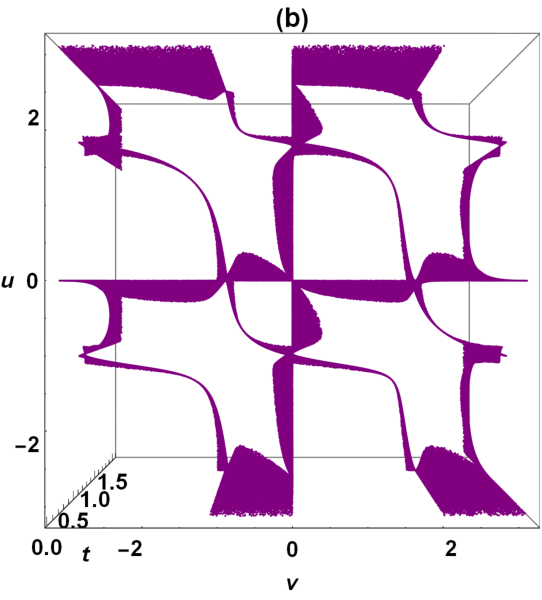}
\end{center}
\caption{Effective classification surface spread by the mapping $(k_x,k_y)
\mapsto (a^4_m,a^6_m,a^4_n,a^6_n)$ with $(m,n)=({-}1,0)$ in (a) and with
$(m,n)=({-}1,1)$ or $(m,n)=(0,1)$ in (b).}
\label{Steff}
\end{figure}
\begin{figure}[H]
\begin{center}
\includegraphics[width=4cm]{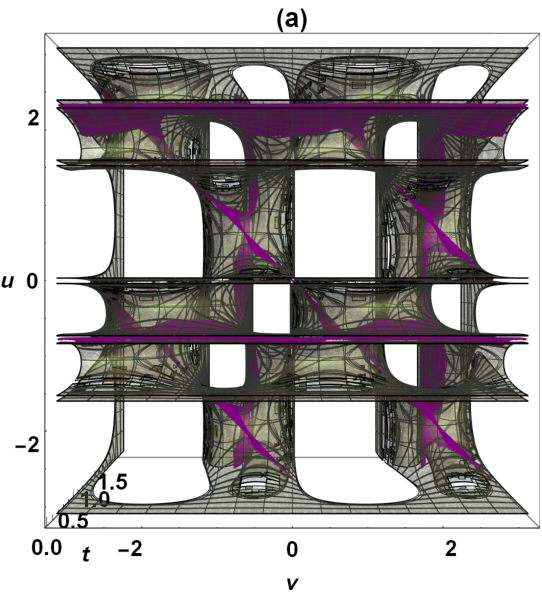}
\includegraphics[width=4cm]{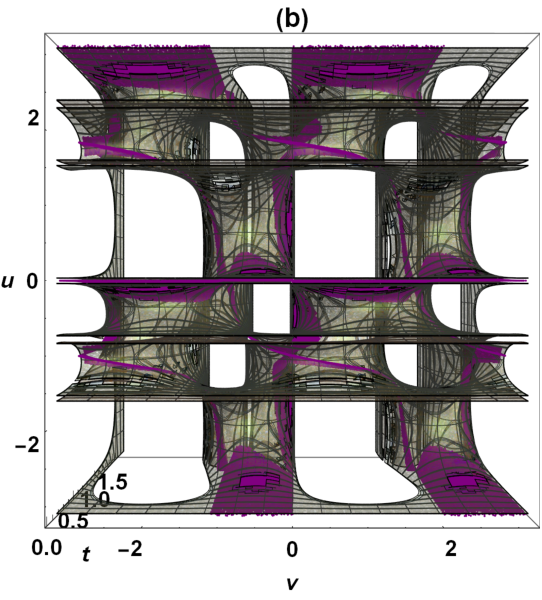}
\end{center}
\caption{View from $t$-direction of how \textit{effective} classification
surface, with $(m,n)=({-}1,0)$ in (a), with $(m,n)=({-}1,1)$ or $(m,n)=(0,1)$ in
(b), spreads into universal classification surface $\cal S$. The holes of 
$\cal S$ in $t$-direction capture all topologically inequivalent paths $\gt$,
revealed in Fig.~\ref{Steff}.}
\label{Stens}
\end{figure}
We do not need to determine the exact topology of $\cal S$ and to investigate
how holes relate one to the other. As shown in Fig.~\ref{Steff}, whatever the
values of $(m,n)$, mapping $(k_x,k_y)\mapsto(a^4_m,a^6_m,a^4_n,a^6_n)$ spreads
over a volume smaller than $\cal S$, which we call \textit{effective}
classification surface. In Fig.~\ref{Stens}, one observes that the 12 holes
inside $\cal S$, parallel to $t$-axis, are enough to characterize the topology
of all paths generated by mapping $\gamma\to\gt$. Indeed, any path turning
around one of these twelve holes cannot retract towards a trivial one, and the
first homotopy group is defined by the number of windings around holes.  We will
henceforth use a schematic view of Fig.~\ref{St}~(a), presented in
Fig.~\ref{stsk}, where the twelve corresponding holes are represented as large
or small circles.

Detailed results are given in section ``\textit{Classification and winding
numbers}'' but we discuss at once the choice of permutation $\{m,n,p\}$.
Let us deal with choices $(m,n)=(0,1)$, $(m,n)=({-}1,1)$ and $(m,n)=({-}1,0)$
(we skip index $p$ which can be deduced), the three others giving identical
results.  With $(m,n)=({-}1,0)$, indices of upper and lower bands, the effective
classification surface is represented on Figs.~\ref{Steff}~(a) and
\ref{Stens}~(a); all paths $\gt$ turn around a hole corresponding to a
\textbf{small} circle in Fig.~\ref{stsk}, which are labelled from $C$ to $J$. If
one of indices $(m,n)$ is 1 --the index of the middle band--, effective
classification surfaces for both choices are represented on
Figs.~\ref{Steff}~(b) and \ref{Stens}~(b); all paths $\gt$ turn around a hole
correspond to a \textbf{large} circle or a \textbf{small} circle (among
$\{D,F,H,J\}$) in Fig.~\ref{stsk}.

From now on, we definitely choose $m={-1}$, $n=0$ (thus $p=1$).  The validity of
the topological classification will be established thanks to the mapping into
$\cal S$, yet one may use easier representations. Therefore, we will now give
two alternatives to mapping $\gamma\to\gt$, defined in surfaces of smaller
dimensions. 
 
\subsection{Bidimensional surface $\widetilde{\cal S}_1$}

This representation is efficient only with $(m,n)=({-}1,0)$ (or
$(m,n)=(0,{-}1)$), not with $m=1$ or $n=1$. Since all expressions are symmetric
or antisymmetric with the exchange $a_m\leftrightarrow a_n$, we keep $m={-}1$
and $n=0$ as for $\cal S$.

\subsubsection{$\widetilde{\cal S}_1$ is a projection of $\cal S$}
\label{Qz}

We construct surface $\widetilde{\cal S}_1$ as a projection of $\cal S$, through
the two separate mappings $\tau_a$ and $\tau_d$:
\begin{eqnarray*}
\tau_a:\quad &(a^4_{\sm1},a^6_{\sm1},a^4_0,a^6_0)&
\mapsto(a^4_{\sm1},a^6_{\sm1})\;,\\
\tau_d:\quad &(a^4_{\sm1},a^6_{\sm1},a^4_0,a^6_0)&\mapsto(a^4_0,a^6_0)\;,
\end{eqnarray*}
where $a$ stands for antidiagonal and $d$ for diagonal, giving \textbf{two
disconnected} bidimensional surfaces. Both have an equal inverted four-leaf
clover shape with four holes (one hole inside each leaf); we call $\cal Q$ the
generic surface, having such shape, which can be described by equation
\begin{eqnarray}
\label{Qapprox}
\forall(a^4)^2+(a^6)^2&\le&\frac34\\
\scriptstyle6\big(\!(a^4)^2+(a^6)^2\!\big)^4+\big(\!(a^4)^2-(a^6)^2\!\big)^4
&\le&
\scriptstyle\frac{205}{43}\big(\!(a^4)^2-(a^6)^2\!\big)^2-\frac13
\big(\!(a^4)^2+(a^6)^2\!\big)^2
\nonumber
\end{eqnarray}
so that $\widetilde{\cal S}_1={\cal Q}_a\times{\cal Q}_d$, where ${\cal Q}_a=
\tau_a({\cal S})$ and ${\cal Q}_d=\tau_d({\cal S})$. Eventually, we will find
that $\tau_a(\gt)$ are non trivial only when $\gamma$ circles a contact point
between lower and middle bands; while $\tau_d(\gt)$ are non trivial only when
$\gamma$ circles a contact point between upper and middle bands, as shown in
Fig.~\ref{surf0}.  $\tau_a\times\tau_d$ is surjective, yet we will see that it
preserves the whole topological classification. We introduce the complex
notation: $z_{1a}=a^6_{\sm1}+\ii\,a^4_{\sm1}$ and $z_{1d}=a^6_0+\ii\,a^4_0$ for
further investigations. 

\subsubsection{Validity}

\begin{figure}[H]
\begin{center}
\includegraphics[width=4cm]{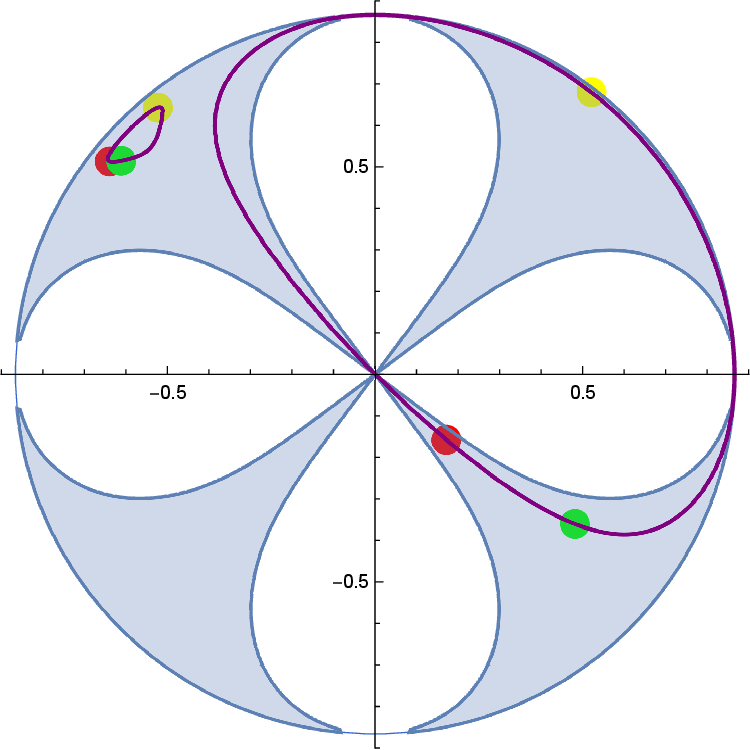}
\end{center}
\caption{Representation of surface ${\cal Q}_d$, in the bidimensional space
spanned by $(a^4_0,a^6_0)$, its outer edge is the circle centered at $O$ (point
$a^4_0=a^6_0=0$) with radius $\frac{\sqrt3}4$, its inner edge is the
symmetric four-leaf clover. The singularity is at $O$. Two loops are shown, with
$t'=\frac12$: a non trivial one, corresponding to a circle around $B_{00}$, of
radius $\frac12$, and a trivial one, corresponding to a circle around $A_{10}$,
of radius $\frac12$. Red, green and yellow points correspond to, respectively,
the beginning, $\frac1{2\pi}$ and $\frac12$ of each path. A similar
representation can be made for ${\cal Q}_a$ \textit{mutas mutantis}, where the
role of $A_{10}$ and $B_{00}$ would be inverted, the representation turned
by $-\frac\pi2$ and the direction of the loop reversed.}
\label{surf0}
\end{figure}

Bloch components $(a^4_m,a^6_m)$ follow (\ref{Qapprox}) for all $t\in[0,1]$ and
$m={-}1$ or $m=0$, but not for $m=1$. Coefficients in (\ref{Qapprox}) are
deduced from a numerical determination of $\widetilde{\cal S}_1$ and must be
improved.  Although their exact determination is still lacking, confidence in
the inverted four-leaf clover shape and in the properties of ${\cal Q}$ is
complete, because our numerical determination is actually exact. ${\cal Q}$ is
embedded in a bidimensional space and reveals a singular point, as shown in
Fig.~\ref{surf0}. Moreover, the outer circular edge is exactly determined
and corresponds to limit $t'=0$, as we shall see.

\subsubsection{Symbolic notations of non trivial loops in
$\widetilde{\cal S}_1$}

Except for some trivial way and return loops, that exhibit two flip points, all
paths $\tau_i(\gt)$ in $\widetilde{\cal S}_1$, with any $i=a,d$, are continuous;
in particular, consider a non trivial path $\tau_i(\gt)$ which goes across the
singularity, $(0,0)\in\tau_i(\gt)$, then $\tau_i(\gt)$ follows necessarily two
consecutive leaves of the clover, therefore it can not retract. In order to
distinguish these path, we will write \coingh, \coingb, \coindh, \coindb,
\coinhg, \coinbg, \coinhd, \coinbd, where each segment stands for the leaf
enveloping it, their intersection for $(0,0)$, and the arrow for the direction
of path $\tau_i(\gt)$. For instance, the path in Fig.~\ref{surf0} writes
$\tau_d(\gt)=$\coinbd. Note that there are other non trivial loops, which we
will study further on.

\subsubsection{Definition of winding number $\omega_1$}

We define $\omega_1$ in $\widetilde{\cal S}_1$ as follow. We use a special
convention, which allows a nice continuation with the winding number defined in
Lieb model.
Consider a loop $\gamma$ in reciprocal space which maps into loops
$\tau_a(\gt)=(a^4_{\sm1}(k_x,k_y),$\penalty-1000$a^6_{\sm1}(k_x,k_y))$ and
$\tau_d(\gt)=(a^4_0(k_x,k_y),$\penalty-10000$a^6_0(k_x,k_y))$. 

We will here only consider non
trivial two-leaf paths as \coingh, \coingb, \coindh, \coindb, \coinhg, \coinbg,
\coinhd\ or \coinbd. Assuming that ${\cal Q}_a$ and ${\cal Q}_d$ are orientated
with their normal pointing in front of Fig.~\ref{surf0}, $\tau_a(\gt)$ and
$\tau_d(\gt)$ are counted positively if they turn in the trigonometric direction
and negatively if they turn in the reverse direction. Then we set $\omega_1$ as
the sum of all windings associated to each path $\tau_a(\gt)$ or $\tau_d(\gt)$
and defined as follow : $\frac14$ for \coingb, \coindh, \coinhg\ or \coinbd; and
${-}\frac14$ for \coingh, \coindb, \coinhd\ or \coinbg. All such paths turn
around singularities $(0,0)$ in ${\cal Q}_a$ or ${\cal Q}_d$, but their non
triviality is proven by the non triviality of $\gt$ in $\cal S$.  This
convention is arbitrary but leads to very convenient connecting rules, in
particular for limits $t'\to0$ and $t'\to1$ and matches further convention of
winding number $\omega_4$.  Other non trivial loops can be observed, that will
be examined further on, but they can be decomposed into these ones, so our
definition of $\omega_1$ is complete.

As will be explained further on, the homotopy classification on
$\widetilde{\cal S}_1$ is equivalent to that on $\cal S$, although $\omega_1$
gives only very partial information. On the contrary, the next four surfaces
${\cal S}_2$, ${\cal R}'$, $\cal C$ and $\cal T$ only provide partial
classification separately. We will define four corresponding winding numbers
$(\omega_2,\omega_3,\omega_4,\omega_5)$ associated to each one.  Eventually, we
will show that topological properties are correctly described when the set
${\cal E}={\cal S}_2\times{\cal R}\times{\cal C}\times\cal T$ (see differences
between $\cal R$ and ${\cal R}'$ afterwards) is used as a space of
classification, with quadruple index $(\omega_2,\omega_3,\omega_4,\omega_5)$.

\subsection{Surface ${\cal S}_2$}

Let us consider ${\cal S}_2$, the first compound in $\cal E$: it is a
bidimensional surface described by equation (\ref{unique}) \textbf{as a function
of} $(X,Y,Z)$, recalling $Z=V+W$.  ${\cal S}_2$ is embedded in a tridimensional
space and reveals a singular point, as shown in Fig.~\ref{surface}.

\subsubsection{${\cal S}_2$ is a projection of $\cal S$}

We construct projection $s$, ${\cal S}\to{\cal S}_2$,
$s(a^4_m,a^6_m,a^4_n,a^6_n)$\penalty-10000 $=\big(X(a^4_m,a^6_m,a^4_n,a^6_n),
Y(a^4_m,a^6_m,a^4_n,a^6_n),Z(a^4_m,a^6_m,a^4_n,a^6_n)\big)$, so ${\cal S}_2=
s({\cal S})$. A loop $\gamma$ in reciprocal space maps into $\gamma_2=s(\gt)$ in
${\cal S}_2$. Consider a path $\gamma$, turning around a contact point, then
$\gamma_2$ turns around $O=(0,0,0)$, the singularity of ${\cal S}_2$; the non
triviality of $\gamma_2$ is proven by that of $\gt$.  Indeed, it is not possible
to reduce $\gamma_2$ continuously without crossing $O$. On the contrary,
$\gamma_2$ (and $\gt$, see however \ref{pointM}) can retract when the surface
delimited by $\gamma$ does not contain any contact point, see
Fig.~\ref{surface}. We choose $m={-}1$ and $n=0$, as already discussed: other
choices prove either equivalent or inefficient. 

\subsubsection{Definition of winding number $\omega_2$}

\begin{figure}[H]
\begin{center}
\includegraphics[width=4cm]{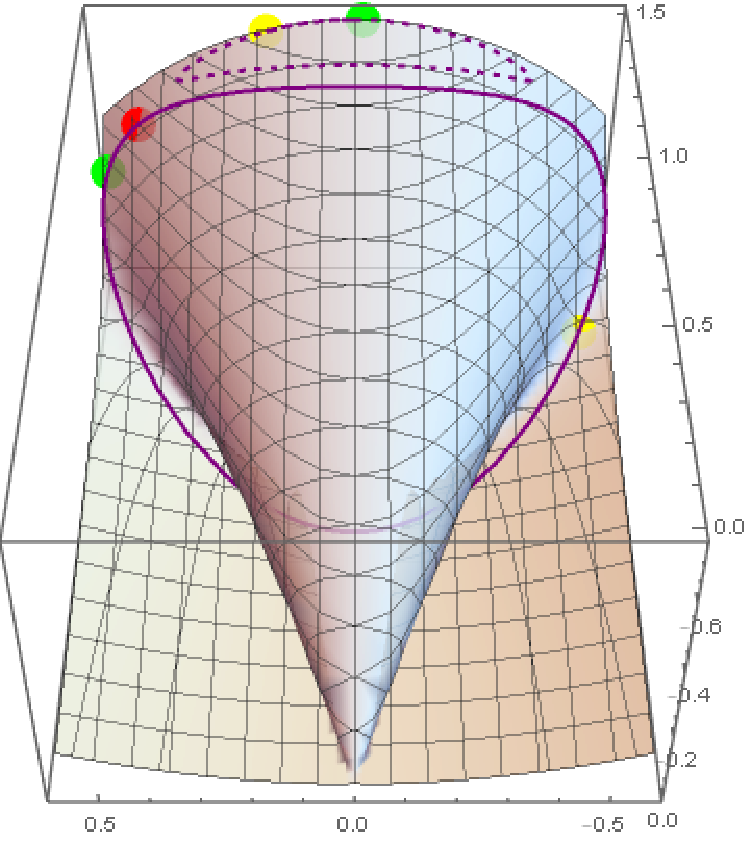}
\end{center}
\caption{Representation of surface ${\cal S}_2$ defined by $3X^2Z+6X^3+3XY^2+Y^4
=0$, where can be observed the singularity at $O=(0,0,0)$.  Using $(X,Y,Z)=
s(a^4_{\sm1}(k_x,k_y),a^6_{\sm1}(k_x,k_y),a^4_0(k_x,k_y),a^6_0(k_x,k_y))$,
the solid line loop corresponds to a circle around point $A_{\sm10}$ with radius
$\frac18$ and reveals non trivial, while the dashed line loop corresponds to a
circle around point $\Gamma_{00}$ with radius $\beta_{1/2}$ and is trivial.  The
cone and the other surface have a common line which coincides with $Z$-axis.
The peak of the cone is $O$, the singularity. Colored points are defined as in
Fig.~\ref{surf0}.}
\label{surface}
\end{figure}

This mapping defines winding number $\omega_2$, which counts algebraically the
number of loops of $\gamma_2$. Surface ${\cal S}_2$ is opened, thus its normal
can be chosen arbitrarily; therefore, $\omega_2$ is defined up to a global sign.
Assuming that ${\cal S}_2$ is orientated with its normal pointing in front of
Fig.~\ref{surface}, $s(\gt)$ is counted positively if it turns in the
trigonometric direction and negatively if it turns in the reverse direction.

\subsubsection{Range of parameters $X$, $Y$ and $Z$}

One observes that parameters $X$ and $Y$ spread over smaller range than what
could be expected from their definition. Indeed,
$X(a^4_m(k_x,k_y),a^6_m(k_x,k_y),a^4_n(k_x,k_y),$
$a^6_n(k_x,k_y))\in[{-}\frac34,0]$
and $Y(a^4_m(k_x,k_y),a^6_m(k_x,k_y),$\penalty-10000 
$a^4_n(k_x,k_y),a^6_n(k_x,k_y))\in
[{-}\frac1{\sqrt3},\frac1{\sqrt3}]$ for all $(k_x,k_y)$, with $m={-}1$ and
$n=0$. This is not true when $m=1$ or $n=1$. In particular, one has
$\forall (k_x,k_y)\in\R^2$,
\begin{equation}
\label{Xneg}
X(a^4_{\sm1}(k_x,k_y),a^6_{\sm1}(k_x,k_y),a^4_0(k_x,k_y),a^6_0(k_x,k_y))\le0\;.
\end{equation}
On the contrary, the range of parameter $Z$ is conform to what can be deduced
from its definition, $\forall(m,n)\in\{{-}1,0,1\}$:
\[
Z(a^4_m(k_x,k_y),a^6_m(k_x,k_y),a^4_n(k_x,k_y),a^6_n(k_x,k_y))\in[0,\frac32]
\]

\subsection{Surface $\cal R$}

We now examine $\cal R$, the second compound in $\cal E$. It is defined through
${\cal R}'$, which is embedded in a bidimensional space and reveals two singular
points, as shown in Fig.~\ref{surf48z2}.

\subsubsection{${\cal R}'$ is a projection of $\cal S$}

\begin{figure}[H]
\begin{center}
\includegraphics[width=4cm]{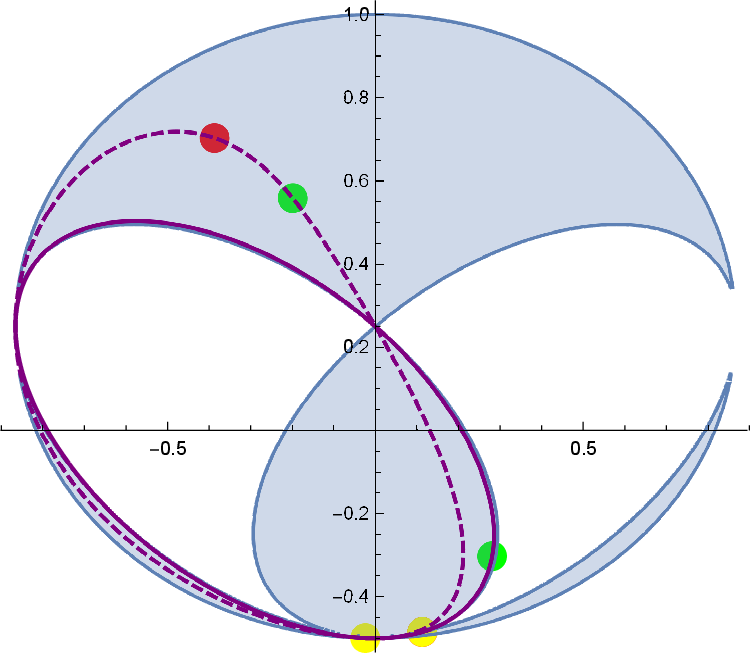}
\end{center}
\caption{Representation of surface ${\cal R}'$, its outer edge is an ellipse,
with semi-major axis along $Ox$ of length $\frac{\sqrt3}2$ and semi-minor axis
along $Oy$ of length $\frac34$, its inner boundary is made of the two ellipses
described in the text. Two non trivial loops are shown, corresponding to circles
with radius $\frac12$ around $B_{00}$ at $t'=\frac12$ (dashed line) or
$\Gamma_{00}$ (solid line) at $t'=1$.  Colored points are defined as in
Fig.~\ref{surf0}, a red point is recovered by a yellow one, which confirms that
the image of the loop around $\Gamma_{00}$ is described twice.}
\label{surf48z2}
\end{figure}

We construct projection $r$ from $\cal S$ to ${\cal R}'$,
$r(a^4_{\sm1},a^6_{\sm1},a^4_0,a^6_0)=(a^4_{\sm1}+a^4_0,a^8_{\sm1}+a^8_0)$, so
${\cal R}'=r({\cal S})$. 
This surface is embedded in an ellipse with semi-major axis along $Ox$ of length
$\frac{\sqrt3}2$ and semi-minor axis along $Oy$ of length $\frac34$, and center
$(0,\frac14)$, see Fig.~\ref{surf48z2}. Two parts are removed, forming
ellipses, inclined by $\pm\frac\pi5$, of equations
\begin{equation}
\label{Rapprox}
\begin{tabular}{l}
$\scriptstyle
\frac94\big((x-\frac1{2\sqrt3})\cos\frac\pi5+y\sin\frac\pi5\big)^2
+
\frac{64}9\big((x-\frac1{2\sqrt3})\sin\frac\pi5-y\cos\frac\pi5\big)^2\le1\;,$
\\
$\scriptstyle
\frac94\big((x+\frac1{2\sqrt3})\cos\frac\pi5-y\sin\frac\pi5\big)^2
+
\frac{64}9\big((x+\frac1{2\sqrt3})\sin\frac\pi5+y\cos\frac\pi5\big)^2
\le1\;,$
\end{tabular}
\end{equation}
except their intersection, which lies inside ${\cal R}'$. Altogether, there are
two holes indeed in ${\cal R}'$.

\subsubsection{Validity}

Confidence in the general shape and in the properties of ${\cal R}'$ is
complete, because our numerical determination is actually exact. However, some
hints indicate that the coefficients in (\ref{Rapprox}) must be improved, see
the discussion about surface $\widetilde{\cal T}$ in appendix. The elliptic
outer edge is exactly determined by equations (\ref{eq1a}), (\ref{eq2a}) and
(\ref{eq3a}) and corresponds to limit $t'=0$. 

\subsubsection{Definition of winding number $\omega_3$}

Assuming that ${\cal R}'$ is orientated with its normal pointing in front of
Fig.~\ref{surf48z2}, $r(\gt)$ is counted positively if it turns in the
trigonometric direction and negatively if it turns in the reverse direction.
Then we set $\omega_3$ as the sum of all windings around any of the two holes.
For instance, a path turning once in the trigonometric direction and enclosing
both holes gives $\omega_3=2$.

$\omega_3$ does not separate a path turning around one hole from that turning
around the other. Let us define a second winding $\widetilde\omega_3$, that
still does not separate holes, but a path around the right hole (such that
horizontal coordinate $x>0$ in Fig.~\ref{surf48z2}) is counted positively, while
a path around the left hole is counted negatively. For instance, a path
enclosing both holes gives $\widetilde\omega_3=0$. 

One can verify that $(\omega_3,\widetilde\omega_3)$ captures the whole topology
of ${\cal R}'$. In appendix, we define quotient spaces $\cal R$ and
$\widetilde{\cal R}$, which are associated to, respectively, winding numbers
$\omega_3$ and $\widetilde\omega_3$. ${\cal R}'$ is homotopically
equivalent to ${\cal R}\times\widetilde{\cal R}$, one finds $\widetilde\omega_3=
-\omega_5$ and the hole, around which $r(\gt)$ turns, is given by $-\omega_3
\omega_5$ with the convention that it is the right hole if $-\omega_3\omega_5>0$
and the left hole if $-\omega_3\omega_5<0$.

\subsection{Surface $\cal C$}

We now examine $\cal C$, the third compound in $\cal E$. $\cal C$ is embedded in
a bidimensional space and reveals two singular points, as shown in
Fig.~\ref{surf13z2}.

\subsubsection{$\cal C$ is a projection of $\cal S$}

We construct projection $c$ from $\cal S$ to $\cal C$,
$c(a^4_{\sm1},a^6_{\sm1},a^4_0,a^6_0)=(a^3_{\sm1}a^1_0+a^1_{\sm1}a^3_0,
a^3_{\sm1}a^3_0-a^1_{\sm1}a^1_0)$, so ${\cal C}=c({\cal S})$. Using complex
notation $z_1$ and ignoring scaling factors $\frac{-Y^2}{2\sqrt{3}(X+V)}$ or
$\frac{-Y^2}{2\sqrt{3}(X+W)}$, it writes $(z_{1a},z_{1d})\mapsto
\overline{(z_{1a}z_{1d})}^2$. One could alternatively study the
surface defined by the mapping\newline
 $(a^4_{\sm1},a^6_{\sm1},a^4_0,a^6_0)\mapsto
(2(a^6_{\sm1}a^4_0+a^4_{\sm1}a^6_0)(a^6_{\sm1}a^6_0-a^4_{\sm1}a^4_0),$%
\penalty-10000
$(a^6_{\sm1}(a^6_0-a^4_0)-a^4_{\sm1}(a^6_0+a^4_0))% 
(a^6_{\sm1}(a^6_0+a^4_0)+a^4_{\sm1}(a^6_0-a^4_0)))$, which corresponds to the
same complex mapping but for the complex conjugation and without any scaling
factor correction.

\begin{figure}[H]
\begin{center}
\includegraphics[width=4cm]{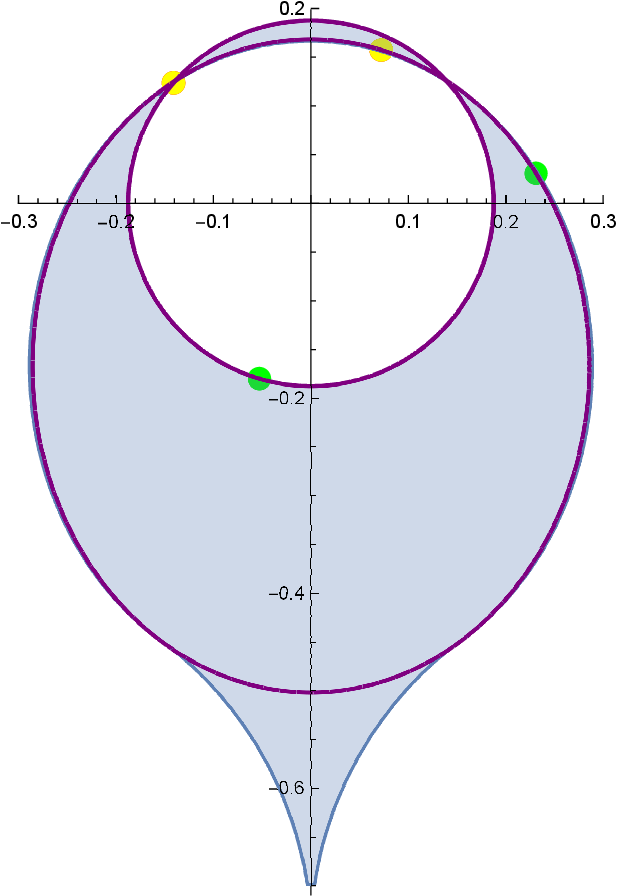}
\end{center}
\caption{Representation of surface $\cal C$, which complicated shape is
described in the text.  Two non trivial loops are shown, one for $t'=1$,
corresponding to a circle around $A_{00}$, with radius $\frac12$, one for
$t'=0$, corresponding to a circle around $M_{00}$, of radius $\frac12$.  Colored
points are defined as in Fig.~\ref{surf0}, both red points are recovered by
yellow ones, which confirms that the image of the loop around $A_{00}$ is
described twice, while that of the loop around $M_{00}$ is described four
times.}
\label{surf13z2}
\end{figure}

$\cal C$ has a unique hole. This hole is delimited by a circle with center
$O=(0,0)$ and radius $\frac{\sqrt3}4$, except for its upper boundary, which is
delimited by the ellipse centered at $(0,{-}\frac16)$, with semi-minor axis
along $Ox$, with length $\frac1{2\sqrt3}$, and semi-major axis along $Oy$, with
length $\frac13$. Above this upper boundary, $\cal C$ is composed of the
crescent of disc; below, by the ellipse; these two parts are connected by two
singular points $(\pm\frac{\sqrt5}{16},\frac18)$. Eventually, at the bottom, the
ellipse is extended by a tail, which we have approximated with two elliptical
arcs, that connect tangentially with the main ellipse, observing that the two
arcs end vertically at $(0,{-}\frac34)$. The whole figure is shown in
Fig.~\ref{surf13z2}.

\subsubsection{Validity}

Most of these parameters are plausible to be exact, the circular boundary is
exactly determined and corresponds to limit $t'=0$, the elliptic one as
well, which corresponds to limit $t'=1$. Thus, the positions of the two
singularities are exact. As for the tail, we have used numerical approximate
values, but one must be aware that, in case the choice of elliptic arcs were
correct, there is only one solution that verifies the given constraints.
Confidence in the general shape is complete, because this surface has been
determined by exact numerical calculations.

\subsubsection{Definition of winding number $\omega_4$}

Assuming that $\cal C$ is orientated with its normal pointing in front of
Fig.~\ref{surf13z2}, $c(\gt)$ is counted positively if it turns in the
trigonometric direction and negatively if it turns in the reverse direction.
Winding number $\omega_4$ is defined by the windings of paths $c(\gt)$ around
the hole in $\cal C$. 

\subsection{Surface $\cal T$}

\begin{figure}[H]
\begin{center}
\includegraphics[width=6cm]{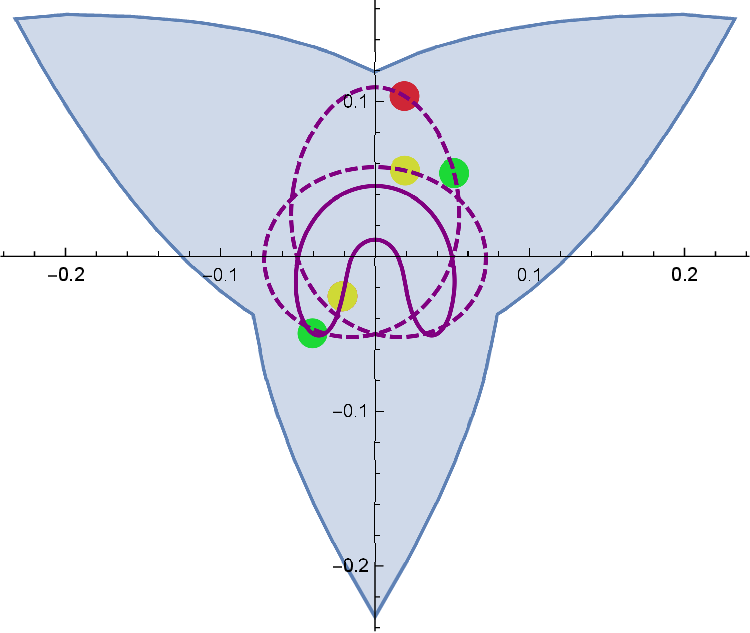}
\end{center}
\caption{Representation of surface $\cal T$, which boundaries are three
congruent arcs of parabolas, the axis of one is vertical while the others are
turned by $\pm\frac{2\pi}3$. Its trilobed shape is very peculiar. A non trivial
path corresponding to a circle around $B_{00}$ with radius $\frac12$ at $t'=1$
is shown (in dashed line), as well as a trivial path (in solid line)
corresponding to a circle around $M_{00}$ with radius 1 at $t'=\frac12$, which
avoids the center. Be aware that the size of $\cal T$ has been rescaled compared
to paths, for convenience. Colored points are defined as in Fig.~\ref{surf0},
the loop at $t'=1$ is described twice but with different paths.}
\label{surf38z2} 
\end{figure}

We now examine $\cal T$, the last compound in $\cal E$. $\cal T$ is embedded in
a bidimensional space and reveals a singular point at $O=(0,0)$. This can not,
however, be proven from its representation in Fig.~\ref{surf38z2} because there
is no hole in $\cal T$. Nevertheless, topological classification is ensured by
the analysis of paths $\gt$ in $\cal S$, so one can use the following results
with full confidence.

\subsubsection{$\cal T$ is a projection of $\cal S$}

We construct projection $t$ from $\cal S$ to $\cal T$,
$t(a^4_{\sm1},a^6_{\sm1},a^4_0,a^6_0)=(a^3_{\sm1}a^8_0+a^3_0a^8_{\sm1},
a^3_{\sm1}a^3_0-a^8_{\sm1}a^8_0)$, so ${\cal T}=t({\cal S})$. Surface $\cal T$
has a beautiful trilobed shape, with no hole but one observes that all paths
$t(\gt)$ avoid the center $O=(0,0)$. More precisely, its boundaries can be
defined by three parabolas of equations $y^2=4x^2-\frac1{16}$ and
$\frac1{16}-y+\sqrt3(\frac1{16}\mp x)= 2\big(\pm
x-\frac1{16}-\sqrt3(y-\frac1{16})\big)^2$.

\subsubsection{Validity}

One of the parabolic boundary of $\cal T$ is exactly determined and corresponds
to limit $t'=0$, which is found analytically. Confidence in the general
shape is complete, since this surface has been found by exact numerical
calculations.

\subsubsection{Definition of winding number $\omega_5$}

Assuming that $\cal T$ is orientated with its normal pointing in front of
Fig.~\ref{surf38z2}, $t(\gt)$ is counted positively if it turns in the
trigonometric direction and negatively if it turns in the reverse direction.
Since all paths $\gt$ turning once in $\cal S$ happen to turn twice in $\cal T$,
$\omega_5$ is defined as half of the windings of paths $t(\gt)$ around
$O=(0,0)$.

\subsection{Reduced topological classification}

We have already established that the topology of Lieb-kagomé energy band
singularities only deals with the way paths $\gt$ are turning around holes $C$
to $J$ in $\cal S$. So eight winding numbers would seem necessary to describe
it. The fundamental group of the effective classification surface reveals,
however, eventually more constrained, since we will find that four winding
numbers, $(\omega_2,\omega_3,\omega_4,\omega_5)$, are enough. As an alternative,
the determination of paths $\tau_a(\gt)$ and $\tau_d(\gt)$ among
\{\coingh,\coingb,\coindh,\coindb, \coinhd,\coinhg,\coinbd,\coinbg\} in
$\widetilde{\cal S}_1$ is enough, although not easy, since it is again more
constrained than what its eight holes seem to indicate.

Altogether, the topological classification surface $\cal S$ can be reduced, in
case $0<t'\le1$, to ${\cal S}_2\times{\cal R}\times{\cal C}\times{\cal T}$,
which fundamental group is $\pi_1({\cal S}_2\times{\cal R}\times{\cal C}
\times{\cal T})=\Z_4$. It can also be determined in $\widetilde{\cal S}_1$, the
fundamental group of which is more complicated but not to be explicitly given.
About the latter, mention should be made of complicated mingling between points
in the circular border of ${\cal Q}_a$ and points in the circular border of
${\cal Q}_d$. We do not need to examine these relations in detail in the general
case, but they prove essential in Lieb limit, $t'=0$, which will be studied
afterwards. Not all classification surfaces are efficient, when extended
formally up to $t'=0$.  Nevertheless, all can be fruitfully used in the vicinity
$t'\sim0$.

\subsection{Topological classification in Lieb case}

Several Dirac points merge at $t'=0$, we will first discuss how to describe
aggregates around points $M_{mn}$.

\subsubsection{Total and partial aggregates around $M_{mn}$ points}
\label{aggregat}

We call $M$-aggregate quadruplet $(A'_{01},B'_{11},B_{00},A_{10})$ and
similar ones translated by $(2\pi m,2\pi n)$ with $(m,n)\in\Z^2$. These points
are close to point $M_{mn}$ and merge altogether at $t'=0$. 

One may also analyse this limit as the merging of partial aggregates. We
therefore define $M_a$-aggregate couple $(A'_{01},A_{10})$ and similar ones
translated by $(2\pi m,2\pi n)$ with $(m,n)\in\Z^2$, $M_d$-aggregates couple
$(B'_{11},B_{00})$ and similar ones translated by $(2\pi m,2\pi n)$ with $(m,n)
\in\Z^2$, $M_l$-aggregates couple $(A'_{01},B_{00})$ and similar ones translated
by $(2\pi m,2\pi n)$ with $(m,n) \in\Z^2$, $M_r$-aggregates couple
$(B'_{11},A_{10})$ and similar ones translated by $(2\pi m,2\pi n)$ with $(m,n)
\in\Z^2$.

Varying parameter $t'$ and using a two-band projection method, it has been
suggested\cite{Lih-King} that each contact points in $M_a$-aggregates have
opposite winding numbers in Lieb limit. We will see that this fits with
$\omega_3$ only. $\forall i\ne3$, windings $\omega_i$ are equal for all contact
points in these aggregates.

We call \textit{single point loop} a path around a single contact point $A$ or
$B$, in reciprocal space. A question arises, when studying limit $t'\to0$:
should the radius of path $\gamma$ change with $t'$?

One finds that, when $t'\to0$, it is necessary to take a radius
$r<(\beta_{t'}-\alpha_{t'})/2
\sim\hspace{-9pt}\raisebox{-4pt}{\tiny0}\;2(t')^3$, in order to
describe a single point loop $\gamma$; otherwise, $\gamma$ contains more than
one singularity.  Therefore, since only paths with none zero radius are to be
considered when $t'=0$, which paths can not be related with single point loops
at $t'>0$, it is not worth considering the latter. 

Lieb limit essentially deals with paths $\gamma$ containing all four contact
points in a $M$-aggregate, because the four merge at $t'=0$.  With $t'\sim0$,
such path $\gamma\mapsto s(\gt)$ trivial in ${\cal S}_2$, $r(\gt)$ trivial in
$\cal R$ and $t(\gt)$ trivial in $\cal T$. On the contrary, $c(\gt)$ is non
trivial in $\cal C$, as shown in Fig.~\ref{surf13z2}, as well as $\tau_a(\gt)$
in ${\cal Q}_a$ or $\tau_d(\gt)$ in ${\cal Q}_d$.  Let us examine this in
details.  We will first study classification surfaces at $t'=0$, then in the
vicinity $t'\sim0$.

\subsubsection{Particular equation in case $t'=0$}

Instead of (\ref{unique}), one observes that Bloch components follow
(\ref{a8b8s4}) for $(m,n)=({-}1,0)$ (or in the reverse order) and (\ref{a8b8s2})
if $m=1$ or $n=1$. However, the latter choice remains non efficient here, so we
will keep the standard choice $m={-}1$ and $n=0$ and consider (\ref{a8b8s4}).

\subsubsection{$\cal S$, ${\cal S}_2$ and ${\cal R}'$ are irrelevant when
$t'=0$}

For $t'=0$, Bloch components follow (\ref{unique}) although the six equations
following it become indeterminate. Using Hopf coordinates, one gets
$r=\rtd$ and $u=v=\frac{3\pi}4$ constant while only $t'$ varies,
whatever path is considered; thus no topological classification can be
performed. Similarly, in ${\cal S}_2$, all coordinates are constant,
$X={-}\frac34$, $Y=0$ and $Z=\frac32$, so $\omega_2$ becomes irrelevant in this
case. Equally, in ${\cal R}'$, all coordinates are constant, $x=0$ and
$y={-}\frac12$, so $\omega_3$ becomes irrelevant in this case.

This is coherent with the triviality of paths $s(\gt)$, $r(\gt)$, while the case
of paths $\gt$ will be examined in the following.

\subsubsection{$\cal T$ and $\cal C$ in case $t'=0$}

When $t'=0$, paths $t(\gt)$ in $\cal T$ are interesting, since they
run along parabola of equation $y^2=4x^2-\frac1{16}$, giving $\omega_5=0$
nevertheless. Paths $c(\gt)$ in $\cal C$ give $\omega_4=\pm4$ (see an example in
Fig.~\ref{surf13z2}).

\subsubsection{Closing of $\widetilde{\cal S}_1$ in case $t'=0$}

$(a^4_{\sm1},a^6_{\sm1},a^4_0,a^6_0)$ follows (\ref{a8b8s4}), thus all paths
$\tau_i(\gt)$ $i=a,d$ lie at the circular boundary of ${\cal Q}_i$, which is
their outer edge. This does not imply that there are no trivial loops in ${\cal
Q}_a$ or ${\cal Q}_d$: trivial loops simply describe a go and back arc, in which
case a discontinuity appears when the direction is changing.

Let us now examine the winding number associated to a circular path at the
border of $\widetilde{\cal S}_1$. If the loop turns once in the trigonometric
direction, it is topologically equivalent to the addition of loops \coinbd\ and
\coinhg, or to the addition of loops \coindh\ and \coingb. If it turns in the
reverse direction, it is topologically equivalent to the addition of loops
\coingh\ and \coindb, or to the addition of loops \coinbg\ and \coinhd. These
combinations appear indeed in the vicinity $t'\sim0$; we do not examine others,
which actually never occur. Eventually, a circular path can be decomposed into
such loops so its winding number is $\frac12=\frac14+\frac14$ or
$-\frac12=-\frac14-\frac14$.

Since $(a^4_{\sm1},a^6_{\sm1},a^4_0,a^6_0)$ follows (\ref{a8b8s4}), there is
only one circular boundary in $\widetilde{\cal S}_1$ when $t'=0$. This follows
from the mingling of ${\cal Q}_a$ and ${\cal Q}_d$. Indeed, one can not
distinguish diagonal or antidiagonal contact points, at $t'=0$, so
$\widetilde{\cal S}_1$ becomes a \textbf{connected closed surface} made of two
disks (with four leave clover shape holes) \textbf{joining} at their
\textbf{mutual circular border}.\footnote{This surface happens to be
homotopically equivalent to Lieb model band structure, which hazard is not
necessary, since we are only interested in the eigenstate space classification.}

Since both $\tau_a(\gt)$ and $\tau_d(\gt)$ describe circle ${\cal S}_1$, the
corresponding winding $\omega_1$ must count double; this will become clear when
studying paths in $\widetilde{\cal S}_1$, in the vicinity $t'\sim0$.

We define ${\cal S}_1$ as the mutual circle with radius $\frac{\sqrt3}2$,
${\cal S}_1$ is homotopically equivalent to $U(1)$. One has $\pi_1(U(1))=\Z$ and
the corresponding winding numbers match exactly $\omega_1$, defined in
$\widetilde{\cal S}_1$. Thus, the classification of Lieb model in
$\widetilde{\cal S}_1$ and in ${\cal S}_1$ are identical.

\subsubsection{Study of $\cal S$ in the vicinity $t'\sim0$}
\label{pointM}

As already stated, a simple way to study Lieb limit in $\cal S$ is to make a
path enclosing points in $M$-aggregates.

All such path $\gamma$ map to path $\gt$, which is the addition of paths
$\gt_i$ around each corresponding hole $i\in\{G,F,D,I\}$, as will be
explained in the next section. However, a special feature appears, depending on
whether $\gt$ encloses $M_{00}$ or not.

\begin{figure}[H]
\begin{center}
\includegraphics[width=4cm]{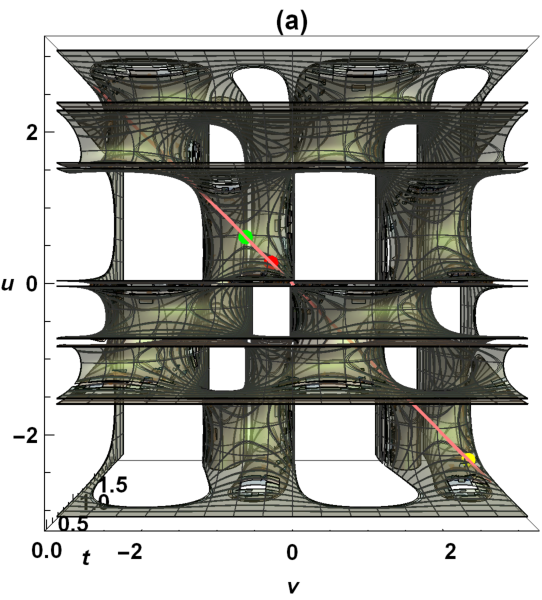}
\includegraphics[width=4cm]{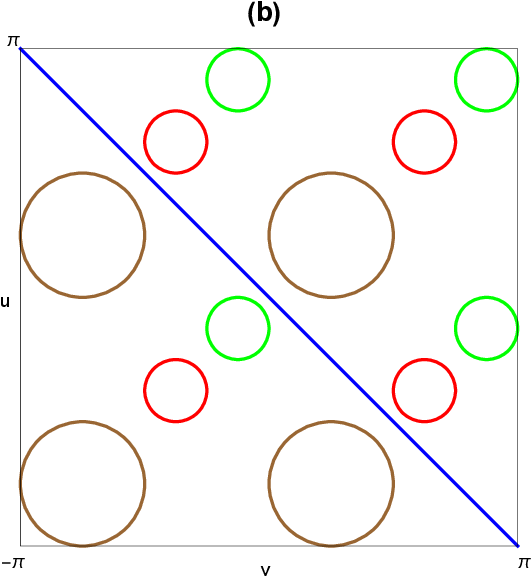}
\end{center}
\caption{Path $\gt_0$ corresponding to $\gamma_0$, a circle around $M_{00}$ with
radius $\frac12$ in surface $\cal S$ (a) or in the schematic representation (b),
with $t'=\frac12$.}
\label{Mnon}
\end{figure}

This feature appears at once when considering a path, called $\gamma_0$, turning
once (in the trigonometric direction) around $M_{00}$ and \textbf{enclosing no}
contact points. $\gamma_0\mapsto\gt_0$, where $\gt_0$ runs across $\cal S$ from
$(u,v)=(-\pi,\pi)$ to $(u,v)=(\pi,-\pi)$ (forgetting about $t$ position, which
is irrelevant here), as shown in Fig.~\ref{Mnon}.

Deciding whether this path is trivial or not is a very intricate question,
because of the complicated folding relations, explained in appendix. However, we
do not need to solve this question, because $M_{mn}$ are not relevant points.
One must subtract path $\gt_0$ from $\gt$, image of any path enclosing $M_{00}$,
thus retrieving the addition of simple paths around holes $C$ to $J$.

\subsubsection{Continuity at $t'=0$}

Mappings $s(\gt)$, $r(\gt)$ and $t(\gt)$ defined in the vicinity\penalty-10000
$t'\sim0$ give results, which are both coherent with the general analysis of
paths at $t'>0$ and that at $t'=0$, as described above. The mappings are
analytical with $t'$ in this vicinity and the homotopy analysis extends
naturally. At $t'>0$, one may consider aggregates, see for instance
$M$-aggregates at $t'=\frac12$ enclosed by a large enough circular loop
$\gamma$, as in Fig.~\ref{surf0prime} where they lead to a non trivial path
(since this surface relates to $\omega_4$) or in Fig.~\ref{surf1}, where they
lead to a trivial path (since this surface relates to $\omega_3$).

\begin{figure}[H]
\begin{center}
\includegraphics[width=5cm]{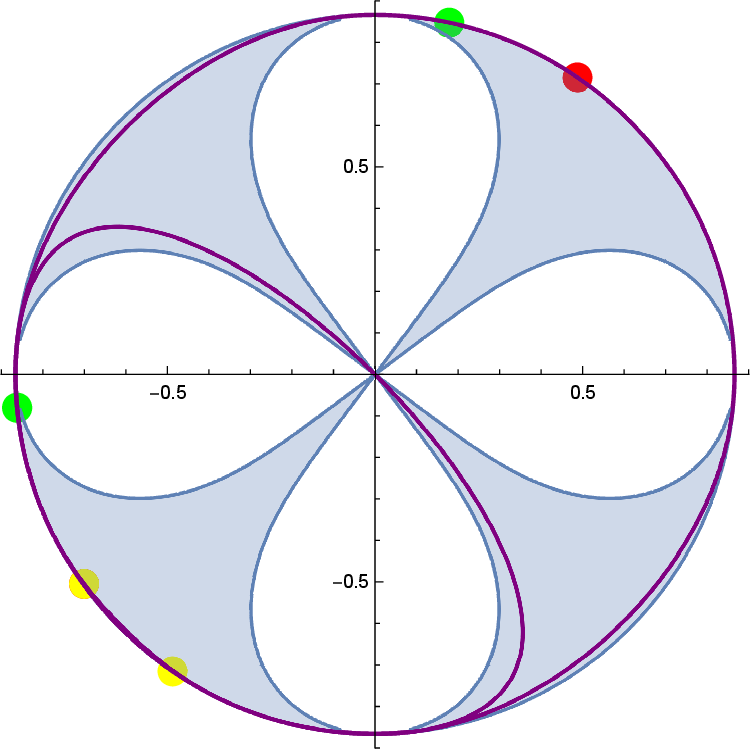}
\end{center} 
\caption{Here are shown two non trivial loops in surface ${\cal Q}_a$: one, for
$t'=\frac12$, corresponding to a circle around $M_{00}$, of radius $\frac52$,
and the other, for $t'=1$, corresponding to a circle around $\Gamma_{00}$, of
radius $\frac52$. Colored points are defined the same way as in
Fig.~\ref{surf0}, a red point is recovered by a yellow one, which confirms that
the image of the loop around $\Gamma_{00}$ is described twice.}
\label{surf0d}
\end{figure}

The study of mappings $\tau_a(\gt)$ and $\tau_d(\gt)$ is more involved. Let us
consider first partial aggregations, defined in subsection \ref{aggregat}.
Paths enclosing $M_l$- and $M_r$-aggregates give separate paths $\tau_a(\gt)$ in
${\cal Q}_a$ and $\tau_d(\gt)$ in ${\cal Q}_d$, which belong to
\{\coingh,\coingb,\coindh,\coindb,\coinhd,\coinhg,\coinbd,\coinbg\}, thus do not
bring any particular light. On the contrary, paths enclosing $M_a$-aggregates
turn around all holes in ${\cal Q}_a$, those enclosing $M_d$-aggregates turn
around all holes inf ${\cal Q}_d$; these paths tend to circular paths at the
outer edge, when $t'\to0$, as shown on Fig.~\ref{surf0d}. If path $\gamma$
encloses both aggregates (that is $M$-aggregate), paths $\tau_a(\gt)$ and
$\tau_d(\gt)$ mingle exactly at $t'=0$ into a simple circular path, which is
however twice degenerate. This explains why the winding, corresponding to both
$M_a$- and $M_d$-aggregates, at $t'=0$, on outer boundary ${\cal S}_1$, must
count double.

Mapping $c(\gt)$ defined in the vicinity $t'\sim0$ is also both coherent with
the general analysis of paths at $t'>0$ and with that at $t'=0$, but the
relation of its corresponding winding number $\omega_4$ with $\omega_1$ must be
discussed here. One observes that any single point loop (defined previously)
turning in the trigonometric direction gives $\omega_4=1$, which corresponds to
$\omega_1=\frac{\omega_4}4$ indeed. Therefore, since path $\gamma$ turning in
the trigonometric direction around any $M$-aggregate is equivalent to four
simple point loops, at $t'>0$, and gives $\omega_4=\pm4$, see for instance
Figs.~\ref{surf0prime} and \ref{surf46z2} which deal with $\omega_4$, though
from another surfaces. This relation extends at $t'=0$ only with the convention
that the circular path, in ${\cal S}_1$, gives $\omega_1=\pm1$, which is the
double from its original definition in ${\cal Q}_a$ or in ${\cal Q}_d$. But,
this double counting has already been explained before and is correct.

\subsubsection{Reduction of the classification surface at $t'=0$}

Whatever surface on which one observes the mapping of paths $\gamma$, in Lieb
limit, one can only define a unique classification index, $\omega_1$.
Altogether, the exact classification space, for Lieb case, is circle
${\cal S}_1$.

\subsection{Topological classification for kagomé case}

Limit $t'\to1$ does not bring as many specificities as the previous one, one
finds that the topological classification in kagomé is identical to that for
$0<t'<1$.  Several Dirac points merge at $t'=1$, we will first discuss how to
describe aggregates around points $\Gamma_{mn}$.

\subsubsection{Aggregates around $\Gamma_{mn}$ points}
\label{aGregat}

We call $\Gamma$-aggregate the couple $(A_{00},A'_{00})$ and similar ones
translated by $(2\pi m,2\pi n)$ with $(m,n)\in\Z^2$. These points are close to
point $\Gamma_{mn}$ and merge altogether at $t'=1$.

Varying parameter $t'$ and using a two-band projection method, it has been
suggested\cite{Lih-King} that each contact points in $\Gamma$-aggregates have
equal winding numbers in kagomé limit. We will see that this fits with all
winding numbers. Kagomé limit deals with paths $\gamma$ containing both contact
points in a $\Gamma$-aggregate.

On the contrary, nothing separates the behavior of paths $\gt$ corresponding to
$\gamma$ around any diagonal contact point $B$ or $B'$ for $t'=1$ from case
$0<t'<1$ (such a path is in Fig.~\ref{surf38z2}); therefore, we do not need to
examine these points at $t'=1$.

\subsubsection{Mapping of contour paths in the vicinity $t'\sim1$}

First of all, consider a simple loop $\gamma$ around contact point
$\Gamma_{00}$, at $t'=1$, then $\gt$ turns twice around hole $C$ in $\cal S$.
The same result is obtained, at $0<t'<1$, if $\gamma$ contains the two
consecutive contact points $A$ and $A'$ in a $\Gamma$-aggregate, that are about
to join.

Similarly, for $0<t'\le1$, $\tau_a(\gt)$ is \coindb, described twice in 
${\cal Q}_a$ (as shown in Fig.~\ref{surf0d}), $s(\gt)$ is also described twice
in ${\cal S}_2$, $r(\gt)$ twice in ${\cal R}'$ (as shown in
Fig.~\ref{surf48z2}), $c(\gt)$ twice in $\cal C$ (as shown in
Fig.~\ref{surf13z2}) and $t(\gt)$ turns four times in $\cal T$. One observes
also double windings in other surfaces, represented in Figs.~\ref{surf0prime},
\ref{surf46z2} and \ref{surf1} and corresponding to windings $\pm\omega_4$ or
$\omega_3$.

Eventually, all the winding numbers depend continuously on $t'$ in kagomé limit. 
\subsubsection{Effective classification surface in kagomé case}

Altogether, the topological classification surface, for kago\-mé case, is
$\cal S$; we have verified that the effective classification surfaces are
exactly that of case $0<t'<1$, which are shown in Fig.~\ref{Steff}.

\section{Classification and winding numbers}

We first focus on the generic situation, for $0<t<1$. We will study Lieb
($t'=0$) or kagomé ($t'=1$) limits afterwards.

\subsection{Case $0<t<1$}

We first study paths $\gt$ in $\cal S$ for each $\gamma$ around a contact point.

\subsubsection{Mapping of paths in $\cal S$}

We consider paths $\gamma$ turning in the trigonometric direction; all
corresponding paths $\gt$ turn once in $\cal S$. We indicate the hole
around which each $\gt$ turns and the direction of $\gt$ (with sign $\pm$) in
Tab.~\ref{tabSurf}.  In the following, we write $\gamma@P$ for ``\textit{path
$\gamma$ turning around object $P$}'' where $P$ can be a point in reciprocal
space or a hole in $\cal S$. We will extend this notation for lists of objects,
like $\gamma@\{P,Q\}$, meaning $\gamma$ turns around $P$ \textbf{or} $Q$. In
Tab.~\ref{tabSurf}, $\gamma@P$ is written vertically for convenience.
\begin{table}[H]
\begin{center}
\begin{tabular}{ccccc}
$\At^\gamma_{A'_{\sm11}}\to
\At^{\gt}_E$&$\At^\gamma_{B'_{01}}\to\At^{\gt}_{\sm H}$&
\qquad&$\At^\gamma_{A'_{01}}\to\At^{\gt}_G$&
$\At^\gamma_{B'_{11}}\to\At^{\gt}_{\sm F}$\\&&&&
\\
\begin{tabular}{c}
$\At^\gamma_{B_{\sm10}}\to\At^{\gt}_{\sm J}$\\
\\
\\
$\At^\gamma_{A'_{\sm10}}\to\At^{\gt}_I$
\end{tabular}&
\multicolumn{3}{c}{\begin{tabular}{|ccc|}
\hline
$\At^\gamma_{A_{00}}\to\At^{\gt}_C$&\qquad&
$\At^\gamma_{B_{00}}\to\At^{\gt}_{\sm D}$\\&&
\\&&\\
$\At^\gamma_{B'_{00}}\to\At^{\gt}_{\sm D}$&\qquad&
$\At^\gamma_{A'_{00}}\to\At^{\gt}_C$\\
\hline
\end{tabular}}&\begin{tabular}{c}
$\At^\gamma_{A_{10}}\to\At^{\gt}_I$\\
\\
\\
$\At^\gamma_{B'_{10}}\to\At^{\gt}_{\sm J}$
\end{tabular}\\&&&&
\\
$\At^\gamma_{B_{\sm1\sm1}}\to\At^{\gt}_{\sm F}$&
$\At^\gamma_{A_{0\sm1}}\to\At^{\gt}_G$&\qquad&
$\At^\gamma_{B_{0\sm1}}\to\At^{\gt}_{\sm H}$&
$\At^\gamma_{A_{1\sm1}}\to\At^{\gt}_E$
\end{tabular}
\end{center}
\caption{Holes around which paths $\gt$ turn and their direction, for any
$0<t'<1$. Black line surrounds ordinary Brillouin zone centered at
$\Gamma_{00}$.}
\label{tabSurf}
\end{table}
One observes that $C$, $E$, $G$, $I$ holes correspond to $A$ or $A'$ points
only, while $D$, $F$, $G$, $J$ holes to $B$ or $B'$ points only.  Each $\gt$,
corresponding to diagonal contact points $B$ or $B'$, turns in the
anti-trigonometric direction; each $\gt$, corresponding to antidiagonal contact
points $A$ or $A'$, turns in the trigonometric direction.  There is no inner
periodicity, so the mapping $\gamma\to\gt$ respects $4\pi$ periodicity in both
$k_x$ and $k_y$ directions.

\subsubsection{Mapping of paths in ${\cal Q}_a$ or ${\cal Q}_d$}

Since all path $\gamma$ around diagonal contact points $B$ or $B'$ give
$\tau_a(\gt)$ trivial in ${\cal Q}_a$, and, equally, all path $\gamma$ around
antidiagonal contact points $A$ or $A'$ give $\tau_d(\gt)$ trivial in
${\cal Q}_d$, we write all non trivial paths in a unique table Tab.~\ref{tabw},
so the reader must understand all paths given for $A$ or $A'$ points as
$\tau_a(\gt)$ and all paths given for $B$ or $B'$ points as $\tau_d(\gt)$. All
$\gamma$ are simple loops in the trigonometric direction.

\begin{table}[H]
\begin{center}
\begin{tabular}{ccccc}
$\At^\gamma_{A'_{\sm11}}\to\,$\coingh&$\At^\gamma_{B'_{01}}\to\,$\coinhd&\qquad&
$\At^\gamma_{A'_{01}}\to\,$\coindh&$\At^\gamma_{B'_{11}}\to\,$\coinhg\\&&&&
\\
\begin{tabular}{c}
$\At^\gamma_{B_{\sm10}}\to\,$\coinbg\\
\\
\\
$\At^\gamma_{A'_{\sm10}}\to\,$\coingb
\end{tabular}&
\multicolumn{3}{c}{\begin{tabular}{|ccc|}
\hline
$\At^\gamma_{A_{00}}\to\,$\coindb&\qquad&$\At^\gamma_{B_{00}}\to\,$\coinbd\\&&
\\&&\\
$\At^\gamma_{B'_{00}}\to\,$\coinbd&\qquad&$\At^\gamma_{A'_{00}}\to\,$\coindb\\
\hline
\end{tabular}}&\begin{tabular}{c}
$\At^\gamma_{A_{10}}\to\,$\coingb\\
\\
\\
$\At^\gamma_{B'_{10}}\to\,$\coinbg
\end{tabular}\\&&&&
\\
$\At^\gamma_{B_{\sm1\sm1}}\to\,$\coinhg&
$\At^\gamma_{A_{0\sm1}}\to\,$\coindh&\qquad&
$\At^\gamma_{B_{0\sm1}}\to\,$\coinhd&$\At^\gamma_{A_{1\sm1}}\to\,$\coingh
\end{tabular}
\end{center}
\caption{Schematic representations of $\tau_d(\gt)$ and $\tau_a(\gt)$ for all
contact points and any $0<t'<1$. Matching rules are coherent with the merging of
Dirac points, both for $t'\to0$ and $t'\to1$.}
\label{tabw}
\end{table}

$A$ and $A'$ contact points are separated from $B$ and $B'$ ones by their
belonging to respectively, ${\cal Q}_a$ or to ${\cal Q}_d$. This separation is
redundantly made from paths $\tau_a(\gt)\in$\{\coindb,\coingh,\coindh,\coingb\}
and $\tau_d(\gt)\in$\{\coinbd,\coinhg,\coinhd,\coinbg\}.

Contact points can merge together only if they fit with apparent matching rules,
that one observes in Tab.~\ref{tabw}: \coindb\ can only match with itself or
with \coingh, and reciprocally; \coindh\ can only match with itself or with
\coingb, and reciprocally; \coinhg\ can only match with itself or with \coinbd,
and reciprocally; \coinhd\ can only match with itself or with \coinbg, and
reciprocally.  These matching rules decorate singularities and will apply for
limits $t'\to0$ and $t'\to1$.

One verifies that $\widetilde{\cal S}_1$ captures all information contained in
$\cal S$. More precisely, the one-to-one relation writes
$C\leftrightarrow$\,\coindb, $D\leftrightarrow$\,\coinbd,
$E\leftrightarrow$\,\coingh, $F\leftrightarrow$\,\coinhg,
$G\leftrightarrow$\,\coindh, $H\leftrightarrow$\,\coinhd, 
$I\leftrightarrow$\,\coingb\ and $J\leftrightarrow$\,\coinbg.

We will verify that $\cal E$ captures all this information, but we must first
detail all winding numbers $\omega_i$, for $i=2,..,5$.

\subsubsection{Winding number $\omega_2$}

Here are the winding numbers $\omega_2(\gamma)$ associated to each $s(\gt)$ in
${\cal S}_2$ for a simple loop $\gamma$ around all contact points $P$, in the
trigonometric direction; we write $\omega_2(P)$ for convenience:
\begingroup
\setlength{\tabcolsep}{3pt}
\begin{table}[H]
\begin{adjustwidth}{-0.5cm}{}
\begin{tabular}{ccccc}
$\omega_2(A'_{\sm11})={-}1$&$\omega_2(B'_{01})={+}1$&\qquad&
$\omega_2(A'_{01})={+}1$&$\omega_2(B'_{11})={-}1$\\&&&&
\\
\begin{tabular}{c}
$\omega_2(B_{\sm10})={+}1$\\
\\
\\
$\omega_2(A'_{\sm10})={+}1$
\end{tabular}&
\multicolumn{3}{c}{\begin{tabular}{|ccc|}
\hline
$\omega_2(A_{00})={-}1$&\qquad&$\omega_2(B_{00})={-}1$\\&&
\\&&\\
$\omega_2(B'_{00})={-}1$&\qquad&$\omega_2(A'_{00})={-}1$\\
\hline
\end{tabular}}&\begin{tabular}{c}
$\omega_2(A_{10})={+}1$\\
\\
\\
$\omega_2(B'_{10})={+}1$
\end{tabular}\\&&&&
\\
$\omega_2(B_{\sm1\sm1})={-}1$&$\omega_2(A_{0\sm1})={+}1$&\qquad&
$\omega_2(B_{0\sm1})={+}1$&$\omega_2(A_{1\sm1})={-}1$
\end{tabular}
\end{adjustwidth}
\caption{$\omega_2(P)$ for all contact points $P$, for any $0<t'<1$.}
\label{tabw2}
\end{table}
$\omega_2$ is constant along diagonals or antidiagonals containing a point
$\Gamma_{mn}$ in reciprocal space. It respects $4\pi$ periodicity along axis
$k_x$ and $k_y$.

It is fruitful to observe how paths $\gt$ dispatch in $\cal S$, depending on the
sign of $\omega_2$. Considering a simple loop $\gamma$ turning around a contact
point, in the trigonometric direction, one finds that $\gt@\{G,H,I,J\}$ if
$\omega_2=1$ and $\gt@\{C,D,E,F\}$ if $\omega_2={-}1$. The first list
corresponds to antidiagonal area in $(u,v)$ space, the second to diagonal area
(see Fig.~\ref{stsk}).

\subsubsection{Winding number $\omega_3$}

Here are the winding numbers $\omega_3(\gamma)$ associated to each $r(\gt)$ in
${\cal R}'$ for a simple loop $\gamma$ around all contact points $P$, in the
trigonometric direction; we write $\omega_3(P)$ for convenience:
\begin{table}[H]
\begin{center}
\begin{tabular}{ccccc}
$\omega_3(A'_{\sm11})={+}1$&$\omega_3(B'_{01})={-}1$&\qquad&
$\omega_3(A'_{01})={-}1$&$\omega_3(B'_{11})={+}1$\\&&&&
\\
\begin{tabular}{c}
$\omega_3(B_{\sm10})={+}1$\\
\\
\\
$\omega_3(A'_{\sm10})={+}1$
\end{tabular}&
\multicolumn{3}{c}{\begin{tabular}{|ccc|}
\hline
$\omega_3(A_{00})={-}1$&\qquad&$\omega_3(B_{00})={-}1$\\&&
\\&&\\
$\omega_3(B'_{00})={-}1$&\qquad&$\omega_3(A'_{00})={-}1$\\
\hline
\end{tabular}}&\begin{tabular}{c}
$\omega_3(A_{10})={+}1$\\
\\
\\
$\omega_3(B'_{10})={+}1$
\end{tabular}\\&&&&
\\
$\omega_3(B_{\sm1\sm1})={+}1$&$\omega_3(A_{0\sm1})={-}1$&\qquad&
$\omega_3(B_{0\sm1})={-}1$&$\omega_3(A_{1\sm1})={+}1$
\end{tabular}
\end{center}
\caption{$\omega_3(P)$ for all contact points $P$, for any $0<t'<1$.}
\label{tabw3}
\end{table}

One observes that $\omega_3$ is constant along approximate\footnote{These lines
become vertical when $t'\to0$ and zigzag when $t'\to1$.}
vertical lines along $k_y$ axis in reciprocal space, more precisely at points
$A_{\mo n}$ and $B'_{\mo n}$ or at points $A'_{\mo n}$ and $B_{\mo n}$, with
$\mo$ fixed and $n$ varying along $\Z$, while it respects $4\pi$ periodicity
along $k_x$ axis.  In $\cal S$, it is $\pi$ periodic along $v$-axis.

It is fruitful to observe how paths $\gt$ dispatch in $\cal S$, depending on the
sign of $\omega_3$. Considering a simple loop $\gamma$ turning around a contact
point, in the trigonometric direction, one finds that $\gt@\{C,D,G,H\}$ if
$\omega_3=1$ and $\gt@\{E,F,I,J\}$ if $\omega_3=-1$. The first list corresponds
to upper area in $(u,v)$ space, the second to lower area.

\subsubsection{Winding number $\omega_4$}

Here are the winding numbers $\omega_4(\gamma)$ associated to each $c(\gt)$ in
$\cal C$ for a simple loop $\gamma$ around all contact points $P$, in the
trigonometric direction; we write $\omega_4(P)$ for convenience:
\begin{table}[H]
\begin{adjustwidth}{-0.5cm}{}
\begin{tabular}{ccccc}
$\omega_4(A'_{\sm11})={-}1$&$\omega_4(B'_{01})={-}1$&\qquad&
$\omega_4(A'_{01})={+}1$&$\omega_4(B'_{11})={+}1$\\&&&&
\\
\begin{tabular}{c}
$\omega_4(B_{\sm10})={-}1$\\
\\
\\
$\omega_4(A'_{\sm10})={+}1$
\end{tabular}&
\multicolumn{3}{c}{\begin{tabular}{|ccc|}
\hline
$\omega_4(A_{00})={-}1$&\qquad&$\omega_4(B_{00})={+}1$\\&&
\\&&\\
$\omega_4(B'_{00})={+}1$&\qquad&$\omega_4(A'_{00})={-}1$\\
\hline
\end{tabular}}&\begin{tabular}{c}
$\omega_4(A_{10})={+}1$\\
\\
\\
$\omega_4(B'_{10})={-}1$
\end{tabular}\\&&&&
\\
$\omega_4(B_{\sm1\sm1})={+}1$&$\omega_4(A_{0\sm1})={+}1$&\qquad&
$\omega_4(B_{0\sm1})={-}1$&$\omega_4(A_{1\sm1})={-}1$
\end{tabular}
\end{adjustwidth}
\caption{$\omega_4(P)$ for all contact points $P$, for any $0<t'<1$.}
\label{tabw4}
\end{table}

$\omega_4=1$ for all four contact points close to $M_{00}$.  $\omega_4={-}1$ for
all four contact points close to $M_{\sm10}$.  $\omega_4$ respects the same
periodicity as $\omega_2$, which allows its complete determination.  $\omega_4$
constant for all contact points close to any point $M_{mn}$ is not surprising
since $\omega_4$ is linked to $\omega_1$, which will operate on these points at
$t'=0$.

It is fruitful to observe how paths $\gt$ dispatch in $\cal S$, depending on the
sign of $\omega_4$. Considering a simple loop $\gamma$ turning around a contact
point, in the trigonometric direction, one finds that the sign of $\gt$ is
$\omega_4$, if $\gt@\{G,H,I,J\}$ and is ${-}\omega_4$ if $\gt@\{C,D,E,F\}$.
Also, $\gt@\{D,F,G,I\}$ if $\omega_4=1$ and $\gt@\{C,E,H,J\}$ if
$\omega_4={-}1$, as represented in Fig.~\ref{cehj}.
\begin{figure}[H]
\begin{center}
\includegraphics[width=3cm]{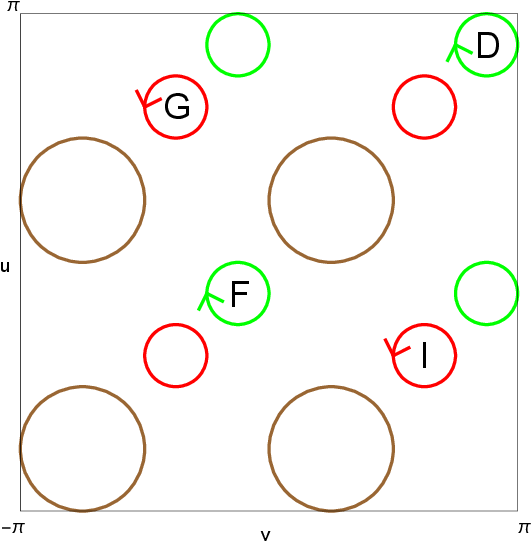}
\end{center}
\caption{Paths $\gt$ corresponding to $\omega_4>0$ in the schematic
representation.}
\label{cehj}
\end{figure}

\subsubsection{Winding number $\omega_5$}

Here are the winding numbers $\omega_5(\gamma)$ associated to each $t(\gt)$ in
$\cal T$ for a simple loop $\gamma$ around all contact points $P$, in the
trigonometric direction; we write $\omega_5(P)$ for convenience:
\begin{table}[H]
\begin{adjustwidth}{-0.5cm}{}
\begin{tabular}{ccccc}
$\omega_5(A'_{\sm11})={+}1$&$\omega_5(B'_{01})={-}1$&\qquad&
$\omega_5(A'_{01})={+}1$&$\omega_5(B'_{11})={-}1$\\&&&&
\\
\begin{tabular}{c}
$\omega_5(B_{\sm10})={-}1$\\
\\
\\
$\omega_5(A'_{\sm10})={+}1$
\end{tabular}&
\multicolumn{3}{c}{\begin{tabular}{|ccc|}
\hline
$\omega_5(A_{00})={+}1$&\qquad&$\omega_5(B_{00})={-}1$\\&&
\\&&\\
$\omega_5(B'_{00})={-}1$&\qquad&$\omega_5(A'_{00})={+}1$\\
\hline
\end{tabular}}&\begin{tabular}{c}
$\omega_5(A_{10})={+}1$\\
\\
\\
$\omega_5(B'_{10})={-}1$
\end{tabular}\\&&&&
\\
$\omega_5(B_{\sm1\sm1})={-}1$&$\omega_5(A_{0\sm1})={+}1$&\qquad&
$\omega_5(B_{0\sm1})={-}1$&$\omega_5(A_{1\sm1})={+}1$
\end{tabular}
\end{adjustwidth}
\caption{$\omega_5(P)$ for all contact points $P$, for any $0<t'<1$.}
\label{tabw5}
\end{table}
$\omega_5$ respects $2\pi$ periodicity in both directions $k_x$ and $k_y$, in
reciprocal space and is constant along any diagonal or antidiagonal.

Considering a simple loop $\gamma$ turning around a contact point, in the
trigonometric direction, one finds that $\omega_5(\gamma)=1$ for all non
diagonal points $A$ or $A'$, while $\omega_5(\gamma)={-}1$ for all diagonal
points $B$ or $B'$. Therefore, $\omega_5=1$ corresponds to $\gt@\{C,E,G,I\}$
(holes associated to $A$ or $A'$ contact points) and $\omega_5={-}1$ to
$\gt@\{D,F,G,J\}$ (those associated to $B$ or $B'$ ones).

\subsubsection{Winding combinations}

Considering a simple loop $\gamma$ turning around a contact point, in the
trigonometric direction, one observes that $\omega_2\omega_4=\omega_5$. If
$\gamma$ turns in the opposite direction, one gets
$\omega_2\omega_4={-}\omega_5$.  Indeed, even combinations are independent of
the direction of $\gamma$, contrary to odd ones.

We have presented this combination $\omega_2\omega_4$ in purpose: indeed,
$\omega_2\omega_4>0$ characterizes non diagonal contact points $A$ or $A'$,
$\omega_2\omega_4<0$ characterizes diagonal ones $B$ or $B'$. Therefore,
$\omega_2\omega_4>0$ characterizes paths $\gamma$ mapping to $\gt@\{C,E,G,I\}$
while $\omega_2\omega_4<0$ characterizes those mapping to $\gt@\{D,F,H,J\}$. Be
aware of the difference with the analysis done with $\omega_5$, which depends on
the direction of $\gamma$ and therefore cannot be conclusive. We write
$\sigma_5=\sign(\omega_2\omega_4)$.

Similarly, $\omega_4\omega_5=\omega_2$ when $\gamma$ turns in the trigonometric
direction, $\omega_4\omega_5={-}\omega_2$ when $\gamma$ turns in the opposite
one.  Thus, $\omega_4\omega_5>0$ characterizes $\gt@\{G,H,I,J\}$ while
$\omega_4\omega_5<0$ characterizes $\gt@\{C,D,E,F\}$. We write
$\sigma_2=\sign(\omega_4\omega_5)$.

Eventually, $\prod_{i=2}^5\omega_i=\omega_3$ when $\gamma$ turns in the
trigonometric direction, $\prod_{i=2}^5\omega_i={-}\omega_3$ when $\gamma$
turns in the opposite one. Thus,  $\prod_{i=2}^5\omega_i>0$ characterizes
$\gt@\{E,F,I,J\}$ while $\prod_{i=2}^5\omega_i<0$ characterizes
$\gt@\{C,D,G,H\}$. We write $\sigma_3=\sign(\omega_2\omega_3\omega_4\omega_5)$.

\subsubsection{Bijection between $\cal E$ and $\cal S$}

Altogether, these combinations allow one to discriminate every hole and recover
all information captured in $\cal S$. One can proceed in two steps. First, the
hole is determinate by $\{\sigma_2,\sigma_3,\sigma_5\}$, as shown in
Tab.~\ref{bilantrous}. Second, once the hole is determinate, the winding
(number of loops and direction) is given by $\omega_5$.
\begin{table}[H]
\begin{center}
\begin{tabular}{l|c|c|c|c|c|c|c|c}
$\cal S$&$C$&$D$&$E$&$F$&$G$&$H$&$I$&$J$\\
\hline
$\sigma_2$&-1&-1&-1&-1&1&1&1&1\\
\hline
$\sigma_3$&1&1&-1&-1&1&1&-1&-1\\
\hline
$\sigma_5$&1&-1&1&-1&1&-1&1&-1\\
\end{tabular}
\end{center}
\label{bilantrous}
\caption{Determination of the hole around which $\gt$ turns.}
\end{table}

The sum of $\omega_i(\gamma)$ for all $\gamma$ making simple loops in the
trigonometric direction around $A'_{01}$, $A_{10}$, $B'_{11}$ and $B_{00}$ is
zero for $i=2,3,5$ and 4 for $i=4$. It actually amounts to the integral of
$\omega_i$ inside a Brillouin zone centered at $M_{00}$. 

The sum of $\omega_i(\gamma)$ for all $\gamma$ making simple loops in the
trigonometric direction around $A'_{00}$, $A_{00}$, $B'_{00}$ and $B_{00}$ is
zero for $i=4,5$ and ${-}4$ for $i=2,3$. It actually amounts to the integral of
$\omega_i$ inside a Brillouin zone centered at $\Gamma_{00}$.

These properties prepare us to limits $t'=0$ and $t'=1$.

\subsection{Particular case $t'=0$}

As already explained, any winding number in Lieb case can be expressed
through $\omega_1$, which can be calculated in ${\cal S}_1$. We do not show a
picture of paths $\tau_a(\gt)$ or $\tau_d(\gt)$ (which are degenerate in this
very case) since they just describe circle ${\cal S}_1$ in the direction
indicated by $\omega_1$.

\subsubsection{Winding number $\omega_1$}

Here, $\gamma$ loops turn around contact points $M_{mn}$, in the trigonometric
direction; we write $\omega_1(P)$ for convenience:
\begin{table}[H]
\begin{center}
\begin{tabular}{ccccc}
\quad&$\omega_1(M_{\sm10})={-}1$&\qquad&
$\omega_1(M_{00})={+}1$&\quad\\
\\
\\
\quad&$\omega_1(M_{\sm1\sm1})={+}1$&\qquad&
$\omega_1(M_{0\sm1})={-}1$&\quad
\end{tabular}
\end{center}
\caption{$\omega_1(P)$ for all contact points $P$, with $t'=0$.}
\label{tab00}
\end{table}

These values fit correctly with the sum of $\omega_4$ windings around each four
contact points merging towards any $M_{mn}$. One observes that $\omega_1$ has
the same periodicity than $\omega_4$ in case $t'>0$.

\subsection{Particular case $t'=1$}

When $t'=1$, there is nothing particular to say about $B$ and $B'$ contact
points, which final positions have been given before; for instance,
$B_{00}=(\frac{2\pi}3,\frac{2\pi}3)$ at $t'=1$. On the contrary, the study of
$A$ and $A'$ contact points is extremely interesting, since they merge by
couples, having equal winding number $\omega_i$, for all $i=2..5$. Using
$\widetilde{\cal S}_1$, one must follow these couples in ${\cal Q}_a$ only, not
${\cal Q}_d$ which is irrelevant here. We examine first the behavior of paths
$\gt$ in $\cal S$.

\subsubsection{Mapping of paths in $\cal S$ with $t'=1$}

Notations are similar to those for case $0<t'<1$, and $\gt@2C$ means that
$\gt$ makes two loops around $C$. All $\gamma$ are simple loops in the
trigonometric direction.
\begin{table}[H]
\begin{center}
\begin{tabular}{ccccccc}
$\ \; \At^\gamma_{\Gamma_{\sm11}}\to\At^{\gt}_{2E}\!\!\!\!\!\!\!\!$&
  &  &$\!\!\!\!\At^\gamma_{\Gamma_{01}}\to\At^{\gt}_{2G}$& & &  \\
 & & $\ \;\At^\gamma_{B'_{01}}\to\At^{\gt}_{\sm H}$& & &
\multicolumn{2}{l}{
\begin{tabular}{cc}
$\!\!\!\!\!\!\!\!\At^\gamma_{B'_{11}}\to\At^{\gt}_{\sm F}$&\end{tabular}}\\
\multicolumn{2}{r}{\begin{tabular}{cc}
&$\At^\gamma_{B_{\sm10}}\to\At^{\gt}_{\sm J}\!\!\!\!\!\!\!\!$\\
$\At^\gamma_{\Gamma_{\sm10}}\to\At^{\gt}_{2I}\!\!\!\!$&\\&
\end{tabular}}&
\multicolumn{3}{c}{\begin{tabular}{|ccc|}
\hline
&&$\At^\gamma_{B_{00}}\to\At^{\gt}_{\sm D}$
\\
&$\At^\gamma_{\Gamma_{00}}\to\At^{\gt}_{2C}$&
\\
$\At^\gamma_{B'_{00}}\to\At^{\gt}_{\sm D}$&&\\
\hline
\end{tabular}}&\multicolumn{2}{l}{\begin{tabular}{cc}&\\
&$\!\!\!\!\!\!\At^\gamma_{\Gamma_{10}}\to\At^{\gt}_{2I}$\\
$\!\!\!\!\!\!\!\!\At^\gamma_{B'_{10}}\to\At^{\gt}_{\sm J}$&
\end{tabular}}\\
\multicolumn{2}{r}{\begin{tabular}{cc}
&$\At^\gamma_{B_{\sm1\sm1}}\to\At^{\gt}_{\sm F}\!\!\!\!\!\!\!\!$ \end{tabular}}
&&&$\!\!\!\!\At^\gamma_{B_{0\sm1}}\to\At^{\gt}_{\sm H}$& & \\
&  & &\ $\At^\gamma_{\Gamma_{0\sm1}}\to\At^{\gt}_{2G}$& & 
\multicolumn{2}{r}{\begin{tabular}{cc}&
\quad$\At^\gamma_{\Gamma_{1\sm1}}\to\At^{\gt}_{2E}$
\end{tabular}}
\end{tabular}
\end{center}
\caption{Holes around which paths $\gt$ turn and their direction, for $t'=1$.}
\label{tab11}
\end{table}
The periodicity observed is identical to that in case $0<t'<1$. All loops around
holes $\{D,F,H,J\}$, corresponding to paths $\gamma$ around points $B$ or $B'$,
are in the reverse direction (as for $0<t'<1$), while all loops around holes
$\{C,E,G,I\}$, corresponding to paths $\gamma$ around points $\Gamma$, turn
twice in the trigonometric direction.

The following table of loops in $\widetilde{\cal S}_1$ may be directly induced
from all previous results.

\subsubsection{Mapping of paths in ${\cal Q}_a$ or ${\cal Q}_d$ with $t'=1$}

Notations are similar to those for case $0<t'<1$ and those of the previous
subsection. In particular, we write all non trivial paths in a unique table
Tab.~\ref{tabq}.  All $\gamma$ are simple loops in the trigonometric direction.

\begin{table}[H]
\begin{center}
\begin{tabular}{ccccccc}
$\ \;\At^\gamma_{\Gamma_{\sm11}}\to2$\coingh&
  &  &$\!\!\!\!\At^\gamma_{\Gamma_{01}}\to2$\coindh& & &  \\
 & & $\ \;\At^\gamma_{B'_{01}}\to\,$\coinhd& & &
\multicolumn{2}{l}{
\begin{tabular}{cc}
$\!\!\!\!\!\!\!\!\At^\gamma_{B'_{11}}\to\,$\coinhg&\end{tabular}}\\
\multicolumn{2}{r}{\begin{tabular}{cc}
&$\!\!\!\!\At^\gamma_{B_{\sm10}}\to\,$\coinbg$\!\!\!\!\!\!\!\!$\\
$\At^\gamma_{\Gamma_{\sm10}}\to2$\coingb&\\&
\end{tabular}}&
\multicolumn{3}{c}{\begin{tabular}{|ccc|}
\hline
&&$\At^\gamma_{B_{00}}\to\,$\coinbd
\\
&$\At^\gamma_{\Gamma_{00}}\to2$\coindb&
\\
$\At^\gamma_{B'_{00}}\to\,$\coinbd&&\\
\hline
\end{tabular}}&\multicolumn{2}{l}{\begin{tabular}{cc}&\\
&$\!\!\!\!\!\!\At^\gamma_{\Gamma_{10}}\to2$\coingb\\
$\!\!\!\!\!\!\!\!\At^\gamma_{B'_{10}}\to\,$\coinbg&
\end{tabular}}\\
\multicolumn{2}{r}{\begin{tabular}{cc}
&$\!\!\!\!\!\!\!\!\At^\gamma_{B_{\sm1\sm1}}\to\,$\coinhg$\!\!\!\!\!\!\!\!$
\end{tabular}}&&&
$\!\!\!\!\At^\gamma_{B_{0\sm1}}\to\,$\coinhd& & \\
&  & &$\At^\gamma_{\Gamma_{0\sm1}}\to2$\coindh& & 
\multicolumn{2}{r}{\begin{tabular}{cc}&
\quad$\At^\gamma_{\Gamma_{1\sm1}}\to2$\coingh
\end{tabular}}
\end{tabular}
\end{center}
\caption{Schematic representations of $\tau_d(\gt)$ and $\tau_a(\gt)$ for all
contact points and $t'=1$.}
\label{tabq}
\end{table}
The bijection between the representation of paths $\gt$ in $\cal S$ and paths
$\tau_a(\gt)$ and $\tau_d(\gt)$ in $\widetilde{\cal S}_1$ is maintained, as well
as the separation between path
$\tau_a(\gt)\in$\{\coindb,\coingh,\coindh,\coingb\}, which are described twice,
and $\tau_d(\gt)\in$\{\coinbd,\coinhg,\coinhd,\coinbg\}.

We will similarly verify that $\cal E$ captures all this information, but we
must first detail all winding numbers $\omega_i$, for $i=2,...,5$.

\subsubsection{Winding numbers $\omega_2$, $\omega_3$, $\omega_4$ and $\omega_5$
for $t'=1$}

Even if they are continuous in the vicinity $t'\sim1$, we must detail winding
numbers in case $t'=1$ because the number and configuration of contact
points is modified.

\begin{table}[H]
\begin{center}
\begin{tabular}{ccccccc}
$\ \;{\omega_2(\Gamma_{\sm11})\atop={-}2}$&
  &  &$\!\!\!\!\!\!\!\!{\omega_2(\Gamma_{01})\atop={+}2}$& & &  \\
 & & $\ \;{\omega_2(B'_{01})\atop={+}1}$& & &
\multicolumn{2}{l}{
\begin{tabular}{cc}$\!\!\!\!\!\!\!\!^{\omega_2(B'_{11})\atop={-}1}$&
\end{tabular}}\\
\multicolumn{2}{r}{\begin{tabular}{cc}
&$\!\!\!\!^{\omega_2(B_{\sm10})\atop={+}1}\!\!\!\!\!\!\!\!$\\
$^{\omega_2(\Gamma_{\sm10})\atop={+}2}$&\\&
\end{tabular}}&
\multicolumn{3}{c}{\begin{tabular}{|ccc|}
\hline
&&$\!\!\!\!^{\omega_2(B_{00})\atop={-}1}$
\\
&$\!\!\!\!^{\omega_2(\Gamma_{00})\atop={-}2}$&
\\
$^{\omega_2(B'_{00})\atop={-}1}$&&\\
\hline
\end{tabular}}&\multicolumn{2}{r}{\begin{tabular}{cc}&\\
&$\!\!\!\!^{\omega_2(\Gamma_{10})\atop={+}2}$\\
$\!\!\!\!\!\!\!\!^{\omega_2(B'_{10})\atop={+}1}$&
\end{tabular}}\\
\multicolumn{2}{r}{\begin{tabular}{cc}
&$^{\omega_2(B_{\sm1\sm1})\atop={-}1}\!\!\!\!\!\!\!\!$\end{tabular}}&&&
$\!\!\!\!\!\!\!\!\!\!\!\!^{\omega_2(B_{0\sm1})\atop={+}1}$& & \\
&  & &$\!\!\!\!^{\omega_2(\Gamma_{0\sm1})\atop={+}2}$& & 
\multicolumn{2}{r}{\begin{tabular}{cc}&
$^{\omega_2(\Gamma_{1\sm1})\atop={-}2}$
\end{tabular}}
\end{tabular}
\end{center}
\vglue-0.5cm
\caption{$\omega_2(P)$ for all contact points $P$, for $t'=1$.}
\label{tab12}
\end{table}
$\omega_2$ is not constant along diagonal or antidiagonal lines in reciprocal 
space, contrary to case $0<t'<1$, because $\Gamma$ points are aligned with
$B$ and $B'$ ones. It respects $4\pi$ periodicity in both directions $k_x$ and
$k_y$.  In $\cal S$, considering a simple loop $\gamma$ turning around a contact
point, in the trigonometric direction, one finds that $\gt@\{G,I\}$ if
$\omega_2=2$, $\gt@\{H,J\}$ if $\omega_2=1$, $\gt@\{C,E\}$ if $\omega_2={-}2$
and $\gt@\{D,F\}$ if $\omega_2={-}1$.

\begin{table}[H]
\begin{center}
\begin{tabular}{ccccccc}
$\ \;^{\omega_3(\Gamma_{\sm11})\atop={+}2}$&
  &  &$\!\!\!\!\!\!\!\!^{\omega_3(\Gamma_{01})\atop={-}2}$& & &  \\
 & & $\ \;^{\omega_3(B'_{01})\atop={-}1}$& & &
\multicolumn{2}{l}{
\begin{tabular}{cc}$\!\!\!\!\!\!\!\!^{\omega_3(B'_{11})\atop={+}1}$&
\end{tabular}}\\
\multicolumn{2}{r}{\begin{tabular}{cc}
&$\!\!\!\!^{\omega_3(B_{\sm10})\atop={+}1}\!\!\!\!\!\!\!\!$\\
$^{\omega_3(\Gamma_{\sm10})\atop={+}2}$&\\&
\end{tabular}}&
\multicolumn{3}{c}{\begin{tabular}{|ccc|}
\hline
&&$\!\!\!\!^{\omega_3(B_{00})\atop={-}1}$
\\
&$\!\!\!\!^{\omega_3(\Gamma_{00})\atop={-}2}$&
\\
$^{\omega_3(B'_{00})\atop={-}1}$&&\\
\hline
\end{tabular}}&\multicolumn{2}{r}{\begin{tabular}{cc}&\\
&$\!\!\!\!^{\omega_3(\Gamma_{10})\atop={+}2}$\\
$\!\!\!\!\!\!\!\!^{\omega_3(B'_{10})\atop={+}1}$&
\end{tabular}}\\
\multicolumn{2}{r}{\begin{tabular}{cc}
&$^{\omega_3(B_{\sm1\sm1})\atop={+}1}\!\!\!\!\!\!\!\!$\end{tabular}}&&&
$\!\!\!\!\!\!\!\!\!\!\!\!^{\omega_3(B_{0\sm1})\atop={-}1}$& & \\
&  & &$\!\!\!\!^{\omega_3(\Gamma_{0\sm1})\atop={-}2}$& & 
\multicolumn{2}{r}{\begin{tabular}{cc}&
$^{\omega_3(\Gamma_{1\sm1})\atop={+}2}$
\end{tabular}}
\end{tabular}
\end{center}
\vglue-0.5cm
\caption{$\omega_3(P)$ for all contact points $P$, for $t'=1$.}
\label{tab13}
\end{table}
$\omega_3$ is constant along verticals in reciprocal space, as in case
$0<t'<1$. It respects $4\pi$ periodicity in directions $k_x$.
In $\cal S$, considering a simple loop $\gamma$ turning around a contact point,
in the trigonometric direction, one finds that $\gt@\{C,G\}$ if $\omega_2=2$,
$\gt@\{D,H\}$ if $\omega_2=1$, $\gt@\{E,I\}$ if $\omega_2={-}2$ and
$\gt@\{F,J\}$ if $\omega_2={-}1$.

\begin{table}[H]
\begin{center}
\begin{tabular}{ccccccc}
$\ \;^{\omega_4(\Gamma_{\sm11})\atop={-}2}$&
  &  &$\!\!\!\!\!\!\!\!^{\omega_4(\Gamma_{01})\atop={+}2}$& & &  \\
 & & $\ \;^{\omega_4(B'_{01})\atop={-}1}$& & &
\multicolumn{2}{l}{
\begin{tabular}{cc}$\!\!\!\!\!\!\!\!^{\omega_4(B'_{11})\atop={+}1}$&
\end{tabular}}\\
\multicolumn{2}{r}{\begin{tabular}{cc}
&$\!\!\!\!^{\omega_4(B_{\sm10})\atop={-}1}\!\!\!\!\!\!\!\!$\\
$^{\omega_4(\Gamma_{\sm10})\atop={+}2}$&\\&
\end{tabular}}&
\multicolumn{3}{c}{\begin{tabular}{|ccc|}
\hline
&&$\!\!\!\!^{\omega_4(B_{00})\atop={+}1}$
\\
&$\!\!\!\!^{\omega_4(\Gamma_{00})\atop={-}2}$&
\\
$^{\omega_4(B'_{00})\atop={+}1}$&&\\
\hline
\end{tabular}}&\multicolumn{2}{r}{\begin{tabular}{cc}&\\
&$\!\!\!\!^{\omega_4(\Gamma_{10})\atop={+}2}$\\
$\!\!\!\!\!\!\!\!^{\omega_4(B'_{10})\atop={-}1}$&
\end{tabular}}\\
\multicolumn{2}{r}{\begin{tabular}{cc}
&$^{\omega_4(B_{\sm1\sm1})\atop={+}1}\!\!\!\!\!\!\!\!$\end{tabular}}&&&
$\!\!\!\!\!\!\!\!\!\!\!\!^{\omega_4(B_{0\sm1})\atop={-}1}$& & \\
&  & &$\!\!\!\!^{\omega_4(\Gamma_{0\sm1})\atop={+}2}$& & 
\multicolumn{2}{r}{\begin{tabular}{cc}&
$^{\omega_4(\Gamma_{1\sm1})\atop={-}2}$
\end{tabular}}
\end{tabular}
\end{center}
\vglue-0.5cm
\caption{$\omega_4(P)$ for all contact points $P$, for $t'=1$.}
\label{tab14}
\end{table}
Non obvious properties of $\omega_4$ in reciprocal space can be observed, which
are not easy to formulate. It respects $4\pi$ periodicity in both directions
$k_x$ and $k_y$. In $\cal S$, considering a simple loop $\gamma$ turning around
a contact point, in the trigonometric direction, one finds that $\gt@\{G,I\}$ if
$\omega_2=2$, $\gt@\{D,F\}$ if $\omega_2=1$, $\gt@\{C,E\}$ if $\omega_2={-}2$
and $\gt@\{H,J\}$ if $\omega_2={-}1$.

\begin{table}[H]
\begin{center}
\begin{tabular}{ccccccc}
$\ \;^{\omega_5(\Gamma_{\sm11})\atop={+}2}$&
  &  &$\!\!\!\!\!\!\!\!^{\omega_5(\Gamma_{01})\atop={+}2}$& & &  \\
 & & $\ \;^{\omega_5(B'_{01})\atop={-}1}$& & &
\multicolumn{2}{l}{
\begin{tabular}{cc}$\!\!\!\!\!\!\!\!^{\omega_5(B'_{11})\atop={-}1}$&
\end{tabular}}\\
\multicolumn{2}{r}{\begin{tabular}{cc}
&$\!\!\!\!^{\omega_5(B_{\sm10})\atop={-}1}\!\!\!\!\!\!\!\!$\\
$^{\omega_5(\Gamma_{\sm10})\atop={+}2}$&\\&
\end{tabular}}&
\multicolumn{3}{c}{\begin{tabular}{|ccc|}
\hline
&&$\!\!\!\!^{\omega_5(B_{00})\atop={-}1}$
\\
&$\!\!\!\!^{\omega_5(\Gamma_{00})\atop={+}2}$&
\\
$^{\omega_5(B'_{00})\atop={-}1}$&&\\
\hline
\end{tabular}}&\multicolumn{2}{r}{\begin{tabular}{cc}&\\
&$\!\!\!\!^{\omega_5(\Gamma_{10})\atop={+}2}$\\
$\!\!\!\!\!\!\!\!^{\omega_5(B'_{10})\atop={-}1}$&
\end{tabular}}\\
\multicolumn{2}{r}{\begin{tabular}{cc}
&$^{\omega_5(B_{\sm1\sm1})\atop={-}1}\!\!\!\!\!\!\!\!$\end{tabular}}&&&
$\!\!\!\!\!\!\!\!\!\!\!\!^{\omega_5(B_{0\sm1})\atop={-}1}$& & \\
&  & &$\!\!\!\!^{\omega_5(\Gamma_{0\sm1})\atop={+}2}$& & 
\multicolumn{2}{r}{\begin{tabular}{cc}&
$^{\omega_5(\Gamma_{1\sm1})\atop={+}2}$
\end{tabular}}
\end{tabular}
\end{center}
\vglue-0.5cm
\caption{$\omega_5(P)$ for all contact points $P$, for $t'=1$.}
\label{tab15}
\end{table}
$\omega_5$ respects $2\pi$ periodicity in both directions $k_x$ and $k_y$, in
reciprocal space, as in case $0<t'<1$.  Considering a simple loop $\gamma$
turning around a contact point, in the trigonometric direction, one observes
that $\omega_5(\gamma)=2$ for all non diagonal points $\Gamma$, while
$\omega_5(\gamma)={-}1$ for all diagonal points $B$ or $B'$. $\omega_5=2$
corresponds to $\gt@\{C,E,G,I\}$ and $\omega_5={-}1$ to $\gt@\{D,F,G,J\}$;
regarding only the sign of $\omega_5$, these correspondences are identical to
that in case $0<t'<1$.
\endgroup

\subsubsection{Bijection $\cal E$ and $\cal S$ when $t'=1$}

Although some winding numbers are modified, the definition of $\sigma_2$,
$\sigma_3$ and $\sigma_5$ is preserved at $t'=1$, thus Tab.~\ref{bilantrous} is
valid and proves that four winding numbers are necessary to describe the
topological properties in this case. Once the hole is determinate, $\omega_5$
gives the direction and number of loops of $\gt$.

\subsection{Merging of contact points}

The analysis of the merging of Dirac contact point when $t'\to1$ comes straight.
Let us use terms defined in subsection \ref{aGregat}. One must consider contact
points in $\Gamma$-aggregates; we have shown that they are homotopically
equivalent, so all windings are equal and remain relevant at $t'=1$. One gets
$\omega_2=\omega_4=2(-1)^{m+n+1}$, $\omega_3=2(-1)^{m+1}$ and $\omega_5=2$
around points $\Gamma_{mn}$. These values follow the periodicities in reciprocal
space that have been given previously. Remembering that $\Gamma_{mn}$ are
contact points between middle and lower energy bands, one observes that the
middle band is parabolic, while the lower band flat, at these points. This is
conform to theoretical predictions\cite{Montambaux} and confirms that our
definitions of winding numbers are correct.

Analysing merging of Dirac contact points when $t'\to0$ can be done with
several interpretations. Let us use terms defined in subsection \ref{aggregat}.
The simpler and more natural way is to consider all contact points in
$M$-aggregates. Only winding $\omega_4$ captures this mechanism: all contact
points in the aggregate have equal winding $(-1)^{m+n}$, where $M_{mn}$ is
the point towards which they merge. The periodicity in reciprocal space is that
of $\omega_1$ and $\omega_4$. Otherwise, one can separate contact points in
$M_a$-aggregates on one side and in $M_d$-aggregates on the other, and consider
the merging of the two corresponding couples. All windings except $\omega_3$
capture this mechanism, $M_a$-aggregates give $\omega_2=2(-1)^{m+n+1}$,
$\omega_4=2(-1)^{m+n}$ and $\omega_5={-}2$, $M_d$-aggregates give
$\omega_2=2(-1)^{m+n}$, $\omega_4=2(-1)^{m+n}$ and $\omega_5=2$. It is not,
however, possible to describe this merging as that of parabolic contact points,
since these partial aggregates only merge at $t'=0$. At last, one can separate
contact points in $M_l$-aggregates on one side and in $M_r$-aggregates on the
other, and consider the merging of the two corresponding couples. Only winding
$\omega_3$ captures this mechanism, $M_l$-aggregates give $\omega_3=
2(-1)^{m+1}$ while, $M_r$-aggregates give $\omega_3=2(-1)^{m}$.

\section{Discussion}

\subsection{Incoherence of vector angle representation}

The classification of topological defects in $\cal S$, representing eigenvectors
as projectors $\Pi_n$, is now achieved. Before these calculations, we have
tried, instead, to use eigenvectors components, which can be expressed through
angular representation and to map paths $\gamma$ in terms of these angles, as
suggested elsewhere.\cite{Lee} However, we have proven that there is, at least,
one mapping showing a discontinuity, which prevents from a complete
determination of singularities. We have discarded the proof, which is too long
already, since we have successfully completed classification by another method,
but we believe it is important to inform of this difficulty.

\subsection{Connection with Gauss-Bonnet theorem}

Gauss-Bonnet theorem allows\cite{Allendoerfer,Fenchel,Chern} one to relate
topological integer numbers, as winding numbers, to the integral over some
closed path $\gamma$ of a physical quantity $\cal A$. This can be done through
Berry connexion.\cite{Berry} One must introduce a pseudo-potential\cite{prl100}
vector ${\cal A}= \langle{\bf k}|\mathbf{\nabla}|{\bf k}\rangle$ and finds
\begin{equation}
\label{gamA}
\int_\gamma{\cal A}d{\bf k}\in\Z.
\end{equation}

In Lieb model, $t'=0$, one finds (using $m=0$ whereas $m={-}1$
would give the same expressions with opposite sign)
\[
H=\frac{\sqrt{2+\cos k_x+\cos k_y}}{\sqrt6}\Big(a^4_0(k_x,k_y)\lambda_1+
a^6_0(k_x,k_y)\lambda_6\Big),
\]
where the Bloch components,\footnote{Note that any two components, among
$(a^1,a^4,a^6)$, could be used, if one is allowed to use permutation symmetry,
as $y\leftrightarrow z$.} in terms of Gell-Mann matrices, relate
directly\cite{Gamel,Goyal} to a pseudo-spin $S$ with $s=1$, thus (\ref{gamA})
becomes\cite{Lih-King}
\[
\int_\gt{\bf n}\frac{d\bf n}{2\pi}\in\Z,
\]
where \textbf{n} is a normalized vector, proportional to $(-a^6,a^4)$,
in the bidimensional Bloch sphere and we have written $\gt$ by similarity,
notwithstanding that $\cal S$ differs from this classification surface. There is
one superfluous degree of freedom; when skipping away this degree of freedom,
one recovers circle ${\cal S}_1$ as an effective classification surface, which
is embedded in the Bloch sphere the same way it is in $\widetilde{\cal S}_1$ at
$t'=0$.

\begin{figure}[H]
\begin{center}
\includegraphics[width=6cm]{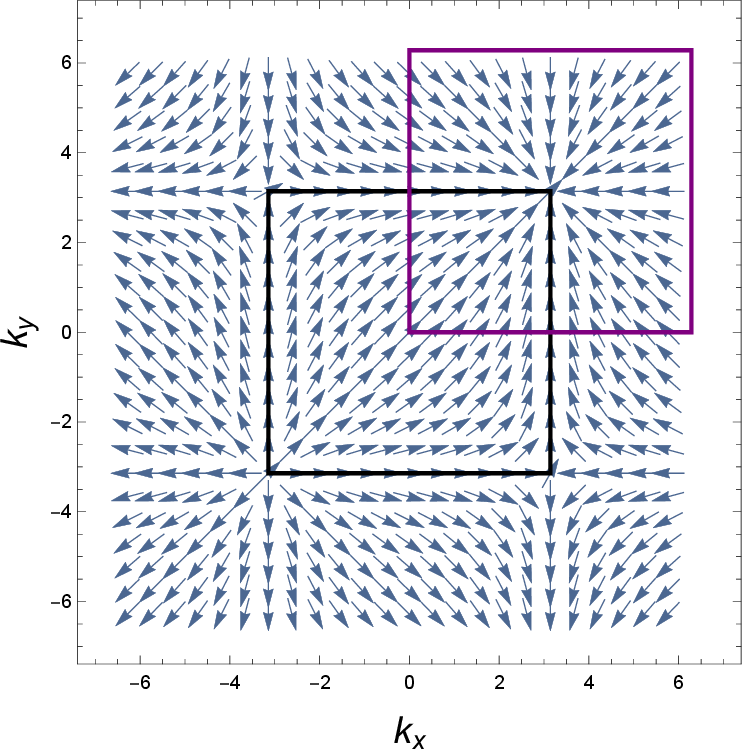}
\end{center}
\caption{Vector field in Lieb model. The winding defect is visible at $M_{00}$,
which arrangement corresponds indeed to $\omega_1=1$.
Two Brillouin zones are indicated, one centered at $\Gamma_{00}$, the other at
$M_{00}$.}
\label{vfLieb}
\end{figure}

One may also use an angular representation $\theta$ such that\cite{Roth}
\[
\int_\gamma\mathbf{\nabla}_{\!\!k}\theta d{\bf k}\in\Z,
\]
this is the reason why the vector components method, described in the previous
subsection, was very tempting, although the interpretation of angles has not
been clarified yet. With an angular representation, the vector field
method\cite{Milnor} applies directly\cite{Montambaux}, as shown on
Fig.~\ref{vfLieb}.

The generalization of this process is very involved in the general case
$0<t'\le1$ and implies integration in $\cal S$. Although (\ref{unique}) is not
homogeneous in $r$, we believe, from our simulations, that the intersection of
$\cal S$ with $r$-$S_3$, the 3-sphere of radius $r$, is independent of $r$ and
could be used as a projective representation of $\cal S$. Also, one must
identify intrinsic angles through Bloch components, in order to apply the vector
field method.

\section{Conclusion}

We have achieved the construction of a topological device for Lieb-kagomé model,
at any $t'\in[0,1]$. Any topologically protected physical state is characterized
by an integer, related to the closed integral of some specific quantity in the
classification surface: this integral is necessarily attached to a winding
integer. 

Protected states of Lieb-kagomé model are defined in reciprocal space as those
at contact points.  In case $0<t'\le1$, on one hand, we have exhibited four
winding numbers $\omega_2$, $\omega_3$, $\omega_4$, $\omega_5$ and proven that
any winding number is proportional to one of these; on the other hand, we have
constructed universal surface $\cal S$, on which any path, measuring a winding
integer, may be mapped and classified.  More precisely, the effective
classification surface fills a volume smaller than $\cal S$	and proves
equivalent to ${\cal E}= {\cal S}_2\times{\cal R}\times{\cal C}\times\cal T$.  
In case $t'=0$, there is only one winding number $\omega_1=\frac{\omega_4}4$ and
the universal surface is ${\cal S}_1$.

Whatever topologically protected state, its properties relate to one, at least,
of these winding numbers and there is a physical quantity, the integral of which
can be performed on the universal classification surface, leading to a complete
characterization of this state. For $0<t'<1$, there are four zero-mass states
per Brillouin zone, associated to each contact point, characterized by winding
number $\omega_i=\pm1$, with $i\in\{2,..,5\}$. For $t'=0$, there is one
zero-mass state per Brillouin zone, characterized by winding number
$\omega_1=\pm1$. For $t'=1$, there are three states per Brillouin zone, two
zero-mass ones associated to winding number $\omega_i=\pm1$, with
$i\in\{2,..,5\}$, and one massive state associated to winding number
$\omega_i=\pm2$, with $i\in\{2,..,5\}$ (this state is protected too).

For Lieb model, the periodicity of $\omega_1$ is completely determinate. Its
sign follows sequence $++--++--\cdots$ in approximate horizontal and vertical
directions. Thus, the effective periodicity is doubled in each direction so the
effective Brillouin zone is four time larger than the real one. In Brillouin
zones centered around $\Gamma_{mn}$, the sign of $\omega_1$ alternates. In
Brillouin zone centered around $M_{mn}$ points, the sign is constant (and equal
to $({-}1)^{m+n}$).  This determination was not possible with previous method
that used both local and two-band equivalence.\cite{Lih-King}

For kagomé model, the periodicity of winding numbers is also determinate, but
depends on which winding is relevant. If $\omega_4$ is relevant, the periodicity
is identical to Lieb case. If $\omega_2$ is relevant, it is almost the same
periodicity, but translated by vector $\overrightarrow{\Gamma_{00}M_{00}}$ so
that $\Gamma_{mn}$ and $M_{mn}$ are inverted in the previous discussion. If
$\omega_3$ is relevant, it is constant along approximate vertical lines, while
it mimics $\omega_2$ periodicity along horizontal ones. Eventually, $\omega_5$
is the only one giving a regular period in each direction: it alternates along
both approximate vertical and horizontal directions and every Brillouin zone is
equivalent.  This determination was not possible with previous method that used
both local and two-band equivalence.\cite{Lih-King}

Lieb case differs from others. Its classification surface ${\cal S}_1$ is unique
and $\omega_1$ corresponds to first homotopy group $\pi_1({\cal S}_1)=\Z$;
while kagomé universal classification surface $\cal S$ contains effective
classification surface ${\cal E}\not=\cal S$; the four winding numbers
$(\omega_2,\omega_3,\omega_4,\omega_5)$ relate to first homotopy group
$\pi_1({\cal E})=\Z^4$ and not to $\pi_1({\cal S})$.

Eventually, this work allows the analysis of the behavior of winding numbers
when $t'$ varies. Taking necessary cautions, all winding numbers depend
\textbf{continuously} on $t'$. When $t'\to1$, winding numbers around diagonal
contact points are unchanged, while those turning around antidiagonal ones
describe two loop paths, because of the merging of singularities. When $t'\to0$,
$\omega_2$, $\omega_3$ and $\omega_5$ become trivial because of the merging of
singularities, while $\omega_4$ describes four loop paths, which explains why
one must described case $t'=0$ with winding number $\omega_1= \frac{\omega_4}4$.

Let us eventually defend the way we have constructed and chosen classification
surfaces. Once we proved that four winding numbers are necessary and sufficient
to describe all paths $\gt$, we tried to collect simple projections, that would
relate to each separate winding number and which fundamental group would be
$\Z$, in order to construct classification surface $\cal E$, which obeys
$\pi_1({\cal E})=\Z^4$. We have almost succeeded, except for $\omega_3$, which
is one of the characteristic numbers of ${\cal R}'$; this classification
surface has two holes and its fundamental group is not simple. Instead, we have
proven that it can be characterized by two integers. We could, however, define
$\cal R$, characterized by $\omega_3$ with $\pi_1({\cal R})=\Z$, the
construction of which has been removed in appendix, because it demands
elaborated mathematical tools.

Classification surface $\widetilde{\cal S}_1$ is given with a different purpose:
as has been explained, its eight holes are sufficient to distinguish all paths
$\gt$. One major interest of $\widetilde{\cal S}_1$ is that ${\cal S}_1$ appears
as a part of $\widetilde{\cal S}_1$ when $t'=0$, which is not even the case for
$\cal S$.  Therefore, $\widetilde{\cal S}_1$ is the only classification surface
valid in the range $0\le t'\le1$, among the three that we have constructed.
Nevertheless, it is indispensable to first build $\cal S$, in order to prove the
non triviality of paths $\gt$.

A prospective study lies in the mapping of angular representation of state
components into $\cal S$; this would let one understand the origin of the
default of this representation. Reminding that the choice of Gell-Mann expansion
of projectors is arbitrary, comparison with other representations\cite{Bloore}
is also very promising.

%\begin{acknowledgements}
\textit{The author thanks deeply Jean-Noël Fuchs for repeated advices and indispensable
remarks during the whole work that have let this article be.}
%\end{acknowledgements}
\appendix

\section{Hamiltonian in basis I}

For completeness,\cite{Lih-King,Parabolic,Bena} let us recall that, in basis I, the Bloch Hamiltonian reads
\[
H_I(\mathbf{k})=
\left(\begin{array}{ccc}0& 1+\e^{\ii k_x}&t'(1+\e^{\ii(k_x+k_y)})\\
1+\e^{-\ii k_x}&0&1+\e^{\ii k_y}\\
t'(1+\e^{-\ii(k_x+k_y)})&1+\e^{-\ii k_y}&0
\end{array}\right).
\]
The relation between both basis is $H(\mathbf{k})=\e^{-\ii \mathbf{k}\cdot
\mathbf{r}}H_S \e^{\ii \mathbf{k}\cdot \mathbf{r}}$ and $H_I(\mathbf{k})=
\e^{-\ii \mathbf{k} \cdot \mathbf{R}}H_S \e^{\ii \mathbf{k} \cdot \mathbf{R}}$,
where $H_S$ is Schrödinger hamiltonian of the crystal and the
complete position operator $\mathbf{r}$ is the sum of the Bravais lattice
position $\mathbf{R}$ and the intra-cell position $\bm{\delta}$. Therefore
\[
H(\mathbf{k})=\e^{-\ii \mathbf{k}\cdot \bm{\delta}}H_I(\mathbf{k}) 
\e^{\ii\mathbf{k}\cdot \bm{\delta}}\;,
\]
with the intra-cell position operator
\[
\bm{\delta} =
\left(\!\!\begin{array}{ccc}-\frac{\mathbf{a}_1}2&0&0 \\
0&0&0\\0&0&\frac{\mathbf{a}_2}2 \end{array}\!\!\right) 
\hbox{ such that } \e^{\ii\mathbf{k}\cdot\bm{\delta}}=
\left(\!\!\begin{array}{ccc}\e^{-\ii k_x/2} &0&0\\
0&1&0\\0&0&\e^{\ii k_y/2} 
\end{array}\!\!\right).
\]

Working with $H_I$ would have the advantage of preserving Brillouin zone, so
that any object be $2\pi$-periodic in $k_x$ and $k_y$ directions. However, $H_I$
is complex, so projectors representing eigenvectors would have eight components,
instead of five with $H$.

\section{Determination of the singularity at the origin of $\cal S$}

First of all, we have defined a supplementary coordinate $s$, in order to get a
bijection from $(a^4_{\sm1},a^6_{\sm1},a^4_0,a^6_0)$ to $(X,Y,Z,s)$.

\paragraph{Definition of coordinate $s$\qquad}

We define $s(a^4,a^6,b^4,b^6)=(a^4)^2+(b^4)^2-(a^6)^2-(b^6)^2$, i.e. $s=V-W$. In
Hopf coordinate, one gets
\begin{eqnarray*}
X&=&\frac{r^2}2(\cos^2t\sin(2u)+\sin^2t\sin(2v))\;;\\
Y&=&\frac{r^2}2\sin(2t)\sin(v-u)\;;\\
Z&=&r^2\;;\\
s&=&r^2\cos(2t)\;.
\end{eqnarray*}

Let us explain why we have needed to introduce this coordinate.

\paragraph{Paths $\gt$ are never close from the singularity}

With \textit{a posteriori} look at paths $\gt$, one observes that the Hopf
coordinate $r=\sqrt{Z}$ comes close to zero for a very short part of the
trajectory and remains close to 1 for the most part of it. We have measured
rigorously the length of $\gt$ in $\cal S$ and found that, indeed, it tends to a
non zero value when the radius of $\gamma$ tends to zero.

\paragraph{Artificial path merging towards the singularity}

In particular, path $s(\gt)$ remains far from $O$, the singularity in
${\cal S}_2$.  With the conviction that approaching the (formerly unsettled)
singularity in $\cal S$ implies approaching $O$ in ${\cal S}_2$ and
understanding that this would never occur with paths $s(\gt)$, we have designed
a path $\gamma_2$ in ${\cal S}_2$ \textbf{which is not a projection} $s(\gt)$.

We have chosen $X=\nu\cos\theta$, $Y=\nu\sin\theta$, $Z$ deduced from
(\ref{unique}) and $s=0$, with $\nu>0$ a parameter. Thanks to bijection
$(X,Y,Z,s)\leftrightarrow(a^4_{\sm1},a^6_{\sm1},a^4_0,a^6_0)$, we could map these
coordinates as path $\gamma_2$ in ${\cal S}_2$ and path $\gt_2$ in $\cal S$.
When $\nu\to0$, the former makes a loop merging towards $O$,
while $\gt_2$ merges towards $(0,0,0,0)$.

This indicates $(0,0,0,0)$ as a possible singularity. We definitively confirmed
this by studying the intersection of $\cal S$ and $r$-$S_3$ spheres, as
explained in the text.

\section{Detailed description of $\cal S$}

We first give the folding rules to which Hopf coordinates $(t,u,v)$ obey ($r$ is
fixed to some arbitrary value and implicit).

\paragraph{Folding rules}

Hopf coordinates are defined with the following folding rules
\begin{eqnarray}
\label{fold1}
(t,u,v)&\leftrightarrow&(-t,u,v\pm\pi)\;,\\
\label{fold2}
(t,u,v)&\leftrightarrow&(\pi-t,u\pm\pi,v)\;,\\
(t,u,v)&\leftrightarrow&(t,u\pm\pi,v\pm\pi)\;,
\label{fold3}
\end{eqnarray}
where $\pm$ depends on $u$ and $v$ and is chosen so that they lie
in the prescribed ranges. 

Within ranges $t\in[0,\frac\pi2[$, $u\in]-\pi,\pi]$ and $v\in]-\pi,\pi]$, these
rules almost never apply. They only apply in the following cases, where we write
intervals following order $(t,u,v)$:

\noindent
(\ref{fold1}) sends $\{\frac\pi2\}\times]{-}\pi,0]\times]{-}\pi,\pi]$ on 
$\{\frac\pi2\}\times]0,\pi]\times]{-}\pi,\pi]$ and reciprocally;

\noindent
(\ref{fold2}) sends $\{0\}\times]{-}\pi,\pi]\times]{-}\pi,0]$ on $\{0\}\times%
]{-}\pi,\pi]\times]0,\pi]$ and reciprocally;

\noindent
(\ref{fold3}) sends $\{0\}\times]{-}\pi,0]\times]{-}\pi,0]$ on
$\{0\}\times]0,\pi]\times]0,\pi]$ and reciprocally;

\noindent
(\ref{fold3}) sends $\{0\}\times]0,\pi]\times]{-}\pi,0]$ on
$\{0\}\times]{-}\pi,0]\times]0,\pi]$ and reciprocally;

\noindent
(\ref{fold3}) sends $\{\frac\pi2\}\times]{-}\pi,0]\times]{-}\pi,0]$ on
$\{\frac\pi2\}\times]0,\pi]\times]0,\pi]$ and reciprocally;

\noindent
(\ref{fold3}) sends $\{\frac\pi2\}\times]0,\pi]\times]{-}\pi,0]$ on
$\{\frac\pi2\}\times]{-}\pi,0]\times]0,\pi]$ and reciprocally.

The $(t,u,v)$ representation of the intersection of $\cal S$ with $r$-$S_3$
(with, for instance, $r=1$) is not only a tridimensional torus with periodic
conditions, it is twisted by these complicated rules. In particular, the
triviality of path $\gt_0$, defined in subsection \ref{pointM}, is left
unsolved.

\paragraph{Detailed descriptions of holes in $\cal S$}

\begin{figure}[H]
\begin{center}
\includegraphics[width=6cm]{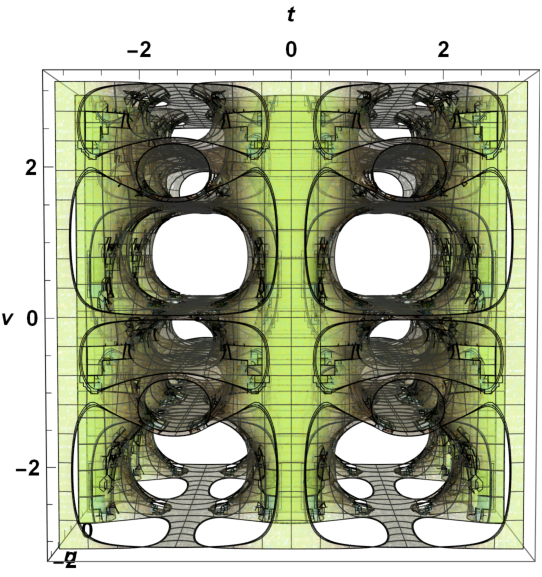}
\end{center}
\caption{View of $\cal S$ in the $u$ direction, with $t$ unfolded twice.}
\label{sixtrous}
\end{figure}

Unfolding $t$ and looking at the intersection of $\cal S$ with $r$-$S_3$ sphere
allows one to better distinguish holes. We show in Fig.~\ref{sixtrous} the view
in the $u$ direction, with $t\in[-\pi,\pi]$, showing two vertical range of six
holes. Taking into accounts folding rules reduces this number to three.

The same view is obtained when $u$ and $v$ are inverted and $t$ is translated
by $\frac\pi2$. This results in a view in the $v$ direction, with
$t\in[-\frac\pi2,\frac{3\pi}2[$.

Altogether, counting the twelve holes in the $t$ direction, one finds exactly 18
holes in $\cal S$.

\section{Definitions of $\cal R$ and $\widetilde{\cal R}$}

\begin{figure}[H]
\begin{center}
\includegraphics[width=4cm]{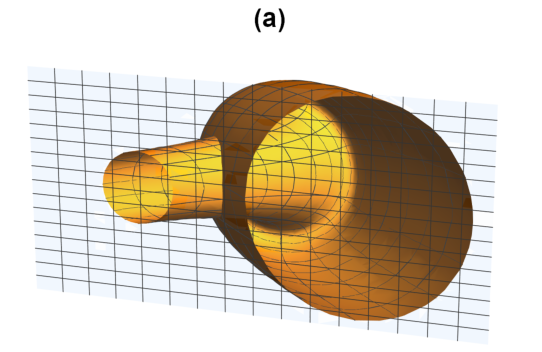}
\includegraphics[width=4cm]{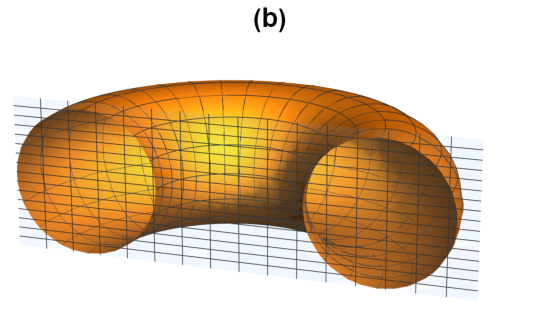}
\end{center}
\caption{Two surfaces related to Klein bottle that allow the construction of
$\cal R$ and $\widetilde{\cal R}$ from ${\cal R}'$. The plane represents surface
${\cal R}'$, where both elliptic holes are deformed in order to match the holes
of the surface --embedded in a tridimensional space-- which lies at the back of
${\cal R}'$.}
\label{klein}
\end{figure}

A simple way to construct $\cal R$ is to adjoin the surface shown in
Fig.~\ref{klein}~(a) to ${\cal R}'$ by matching the two circular edges to each
elliptic hole boundary: one must match the hole boundary to one of the circular
edge of Fig.~\ref{klein}~(a) and the other hole boundary to the other circular
edge, by deforming them.

A simple way to construct $\widetilde{\cal R}$ is to adjoin the surface shown in
Fig.~\ref{klein}~(b) to ${\cal R}'$ by matching the two circular edges to each
elliptic hole boundary: one must match the hole boundary to one of the circular
edge of Fig.~\ref{klein}~(b) and the other hole boundary to the other circular
edge, by deforming them.

Instead, one can construct $\cal R$ and $\widetilde{\cal R}$ as quotient spaces,
using equivalence relations that identify in ${\cal R}'$ paths $r(\gt)$ around
left and right holes in, respectively, the same or the reverse direction. 

\section{Other winding numbers}

Many classification surfaces can be built.  Note that all combinations of
winding integers can be obtained. For instance, mapping $z\mapsto z^2$ can be
applied to $\tau_d(\gt)$, giving $z_{1d}^2=
(a^6_0)^2-(a^4_0)^2+2\,\ii\,a^6_0a^4_0$, and leads to a surface, which is
characterized by the same winding $\omega_4$.

In some of them, only diagonal contact points give non zero winding numbers, or
reversely, only antidiagonal ones, in some of them all contact points are
concerned. Some have two disconnected compounds, while others are connected.
They are all projections from $\cal S$, thus corresponding winding numbers are
product combinations of $\omega_i$, with $i=2..5$.

\paragraph{Alternative bidimensional surface ${\cal S}'_2$\qquad}

We introduce $\tilde Z=\frac34-Z-2X$. Then (\ref{unique}) is equivalent to
\begin{equation}
\label{intermediaire}
3X^2\tilde Z=(Y^2+\frac32X)^2\ ,
\end{equation}
where the dependency in $(a^4,a^6,b^4,b^6)$ is hidden. 
\begin{figure}[H]
\begin{center}
\includegraphics[width=4cm]{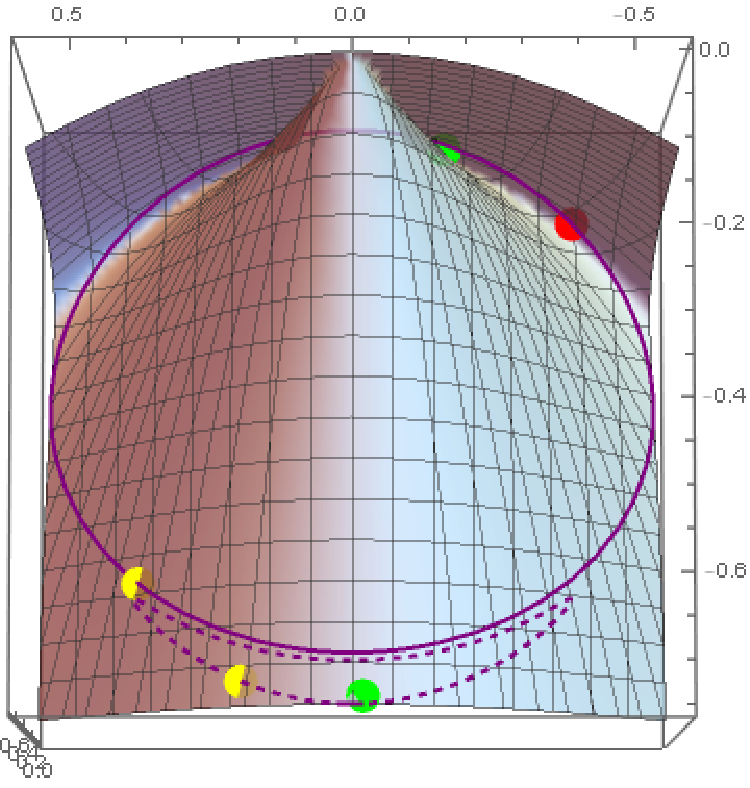}
\end{center}
\caption{Representation of surface ${\cal S}'_2$ defined by $3X^2\tilde
Z=(Y^2+\frac32X)^2$, showing a singularity at $P_0$. Images of the same loops as
those defined in Fig.~\ref{surface} are shown here.  Contact lines are exactly
parallel to $\tilde Z$-axis and contains singularity $P_0$.}
\label{surfbis}
\end{figure}

From $X\in[{-}\frac34,0]$, one get $\tilde Z\in[0,\frac34]$. Then, taking
advantage of these conditions, (\ref{intermediaire}) is equivalent to
\begin{equation}
\label{bis}
\sqrt3({-}X)(\frac{\sqrt3}2\pm\sqrt{\tilde Z})=Y^2\;.
\end{equation}

We note $P_0=(0,0,\frac34)$ the singular point.  (\ref{intermediaire}) or
(\ref{bis}) define new surface ${\cal S}'_2$ (see Fig.~\ref{surfbis}), which is
topologically identical to ${\cal S}_2$.  ${\cal S}'_2$ is embedded in a
tridimensional space and reveals singular point $P_0$, as shown in
Fig.~\ref{surfbis}.  All sections orthogonal to $YP_0\tilde Z$ axis or to
$XP_0Y$ axis are delimited by parabolas.  The mapping of paths $\gamma$ onto
${\cal S}'_2$ is characterized by winding number $\omega_2$.

\paragraph{Alternative surface $\widetilde{\cal C}$}

We define $\widetilde{\cal C}$ the bidimensional surface, built as the
disjunctive union of the two ellipses of equations $\forall(a^1,a^3)\in\R^2$,
\begin{equation}
\begin{tabular}{rcl}
$3\Big(a^1-\frac1{2\sqrt3}\Big)^2+4(a^3)^2$&$\le$&$1\;;$\\
$3\Big(a^1+\frac1{2\sqrt3}\Big)^2+4(a^3)^2$&$\le$&$1\;.$
\end{tabular}
\label{S0prime}
\end{equation}
Note that $\widetilde{\cal C}$ is included in the disk centered at
$(a^1,a^3)=(0,0)$ and with radius $\frac{\sqrt3}2$, which one deduces from
equations (\ref{eq1a}), (\ref{eq2a}), (\ref{eq3a}) (and actually from equation
(\ref{eq1b}), which they follow).

All Bloch components $(a^1_m,a^3_m)$ follow (\ref{S0prime}) for all $t\in[0,1]$
and $m={-}1$ or $m=0$, but not for $m=1$. We construct two disconnected surfaces
as projections of $\cal S$. The mappings write $(a^4_{\sm1},a^6_{\sm1},a^4_0,a^6_0)
\mapsto(a^1_m,a^3_m)$.  With $m={-}1$, mapping $\gt\to(a^1_{\sm1},a^3_{\sm1})$
defines surface $\widetilde{C}_a$; With $m=0$, mapping $\gt\to(a^1_0,a^3_0)$
defines surface $\widetilde{C}_d$.  Paths on $\widetilde{\cal C}_a$ are non
trivial only if $\gamma$ circles a contact point between lower and middle bands.
Paths on $\widetilde{\cal C}_d$ are non trivial only if $\gamma$ circles a
contact point between upper and middle bands. 

Coefficients in (\ref{S0prime}) are deduced from a numerical determination of
$\widetilde{\cal C}$. The intersection of the edges of the two ellipses,
computed with (\ref{S0prime}), are found to be $S_\pm=(0,\pm\frac{\sqrt3}4)$,
which matches exactly limit $t'=0$, where paths are circular, with
radius $\frac{\sqrt3}4$. In addition, the simplicity of these equations and
their coefficients make them very plausible.

\begin{figure}[H]
\begin{center}
\includegraphics[width=5cm]{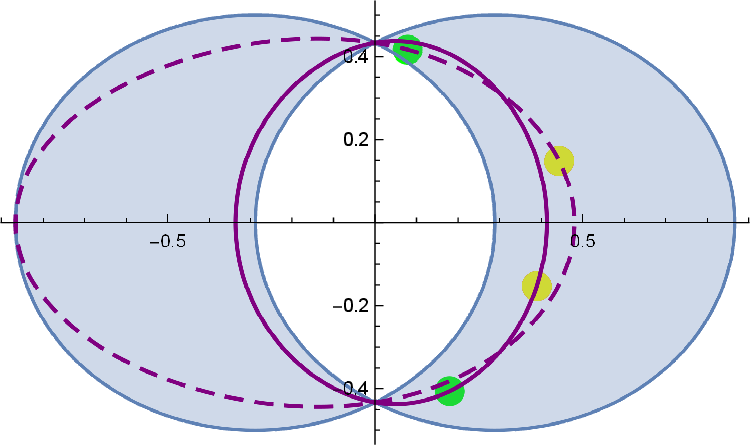}
\end{center}
\caption{Representation of surface $\widetilde{\cal C}_d$, in the bidimensional
space spanned by $(a^1_{\sm1},a^3_{\sm1})$, it is exactly the union of two ellipses
less their intersection. The edges of the two ellipses intersect at points $S_+$
and $S_-$, defined in the text, which are the singularities of the surface.  We
show a path for $t'=\frac12$, corresponding to a circle around $M_{00}$, of
radius $\frac52$ (in dashed line),  and another, for $t'=1$, corresponding to a
circle around $\Gamma_{00}$, of radius $\frac52$ (in solid line).  Colored
points are defined the same way as in Fig.~\ref{surf0}, both red points are
recovered by yellow ones, which confirms that the image of the path around
$\Gamma_{00}$ is described twice, while that of the loop, which contains all
contact points of a $M$-aggregate, is described four times (as would be that of
a path around $M_{00}$ at $t'=0$).}
\label{surf0prime}
\end{figure}

Although we have not proven it yet, confidence in the two-ellipse shape and in
the properties of $\widetilde{\cal C}$ is complete, because our numerical
determination is actually exact. $\widetilde{\cal C}$ is embedded in a
bidimensional space and reveals two singular points, which are the intersection
$S_\pm$, as shown in Fig.~\ref{surf0prime}.

When mingling both surfaces, paths are characterized by winding number
$-\omega_4$. On the contrary, encoding by 1 paths in $\widetilde{\cal C}_a$ and
by ${-}1$ those in $\widetilde{\cal C}_d$ exactly corresponds to
$\omega_5$.  Altogether, $\widetilde{\cal C}$ is globally characterized
by $(-\omega_4,\omega_5)$. Eventually, one finds that the path is non trivial in
$\widetilde{\cal C}_a$ if $-\omega_4\omega_5>0$ and non trivial in 
$\widetilde{\cal C}_d$ if $-\omega_4\omega_5<0$.

\paragraph{Relation between $\widetilde{\cal C}_a\times\widetilde{\cal C}_d$
and $\widetilde{\cal S}_1$\qquad}

Ignoring scaling factors $\frac{-Y^2}{2\sqrt{3}(X+V)}$ or
$\frac{-Y^2}{2\sqrt{3}(X+W)}$, one observes that mapping ${\cal Q}_i\to
\widetilde{\cal C}_i$ corresponds to the conformal mapping $(x,y)\mapsto
(x^2-y^2,-2xy)$, where $x$ stands for $a^6_m$, $y$ for $a^4_m$, $x^2-y^2$ for
$a^3_m$ and $-2xy$ for $a^1_m$ (both last components must be divided by the
scaling factor), with $m={-}1$ when $i=a$ and $m=0$ when $i=d$. Using complex
notation $z_{1i}0$, introduced in \ref{Qz}, these mappings write
$z_{0i}\mapsto\overline{(z_{0i})}^2$, for $i=a,d$.

\paragraph{Alternative surface ${\cal C}'$\qquad\qquad}

We construct surface ${\cal C}'$ as a projection of $\widetilde{\cal S}_1$ by
the mapping
$(a^4_{\sm1},a^6_{\sm1},a^4_0,a^6_0)\mapsto(a^6_{\sm1}a^4_0+a^6_0a^4_{\sm1},
a^6_{\sm1}a^6_0-a^4_{\sm1}a^4_0$. Using the complex notation $z_{1i}$, with
$i=a,d$, it writes $(z_{0a},z_{0d})\mapsto z_{0a}z_{0d}$.

\begin{figure}[H]
\begin{center}
\includegraphics[width=4cm]{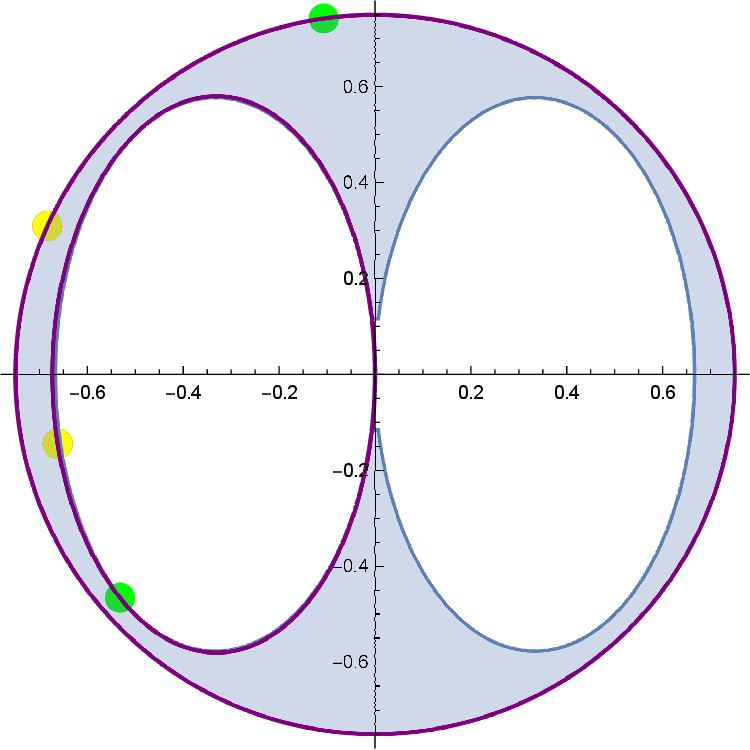}
\end{center}
\caption{Representation of surface ${\cal C}'$, its outer edge is the circle
centered at $O$, with radius $\frac34$, its inner boundary is made of the two
ellipses, described in the text.  Images of the same non trivial loops as in
Fig.~\ref{surf13z2} are shown. Colored points are defined the same way as in
Fig.~\ref{surf0}, both red points are recovered by yellow ones, but here the
images of both loops are described twice.}
\label{surf46z2}
\end{figure}

This surface is a circle with center $O=(0,0)$ and radius $\frac34$, less two
disconnected elliptic holes, shown in Fig.~\ref{surf46z2}; the two ellipses are
centered at $(\pm\frac13,0)$, their semi-minor axis is along $Ox$ with length
$\frac1{\sqrt3}$, their semi-major axis is along $Oy$ with length $\frac13$.

The parameters of ${\cal C}'$ are very likely to be exact, in particular, its
circular edge can be proven. The confidence in the general shape is complete,
since this surface has been found by exact numerical calculations.

One can build two winding numbers the same way we have done for ${\cal R}'$.
Mingling the two holes, one gets $\omega_4$. Doing as for $\widetilde{\cal R}$,
one gets $-\omega_5$.  Altogether, ${\cal C}'$ is globally characterized by
$(\omega_4,-\omega_5)$, as $\widetilde{\cal C}$. Eventually, one finds that the
hole, around which the path turns, is given by $-\omega_4\omega_5$ with the
same convention as for ${\cal R}'$.

\paragraph{Bidimensional surface $\widetilde{\cal T}$}

We write $\widetilde{\cal T}$ the bidimensional surface, defined as the
disjunctive union of the two ellipses of equations
\begin{eqnarray}
\nonumber
\!\!\!\!\!
\begin{tabular}{l}
$\scriptstyle%
\frac94\big(\cos\frac\pi5(a^4-\frac1{2\sqrt3})-\sin\frac\pi5\;a^8\big)^2
+
\frac{64}9\big(\sin\frac\pi5(a^4-\frac1{2\sqrt3})+\cos\frac\pi5\;a^8\big)^2
\le1\;$;
\\
$\scriptstyle%
\frac94\big(\cos\frac\pi5(a^4+\frac1{2\sqrt3})+\sin\frac\pi5\;a^8\big)^2
+
\frac{64}9\big(-\sin\frac\pi5(a^4+\frac1{2\sqrt3})+\cos\frac\pi5\;a^8\big)^2
\le1\;$;
\end{tabular}
%\\
\label{S1approx}
\end{eqnarray}
these ellipses have symmetric axis, turned by $\pm\frac\pi5$ from the $a^4$
axis.

All Bloch components $(a^4_m,a^8_m)$ follow (\ref{S1approx}) for all $t\in[0,1]$
and $m={-}1$ or $m=0$, but not for $m=1$. We construct two disconnected surfaces
as projections of $\cal S$. The mappings write $(a^4_{\sm1},a^6_{\sm1},a^4_0,a^6_0)
\mapsto(a^4_m,a^8_m)$. With $m={-}1$, mapping $\gt\to(a^4_{\sm1},a^8_{\sm1})$
defines surface $\widetilde{\cal T}_a$; with $m=0$, mapping $\gt\to
(a^4_0,a^8_0)$ defines surface $\widetilde{\cal T}_d$.  $\widetilde{\cal T}_a$
are non trivial only if $\gamma$ circles a contact point between lower and
middle bands.  Paths on $\widetilde{\cal T}_d$ are non trivial only if $\gamma$
circles a contact point between upper and middle bands. 

\begin{figure}[H]
\begin{center}
\includegraphics[width=5cm]{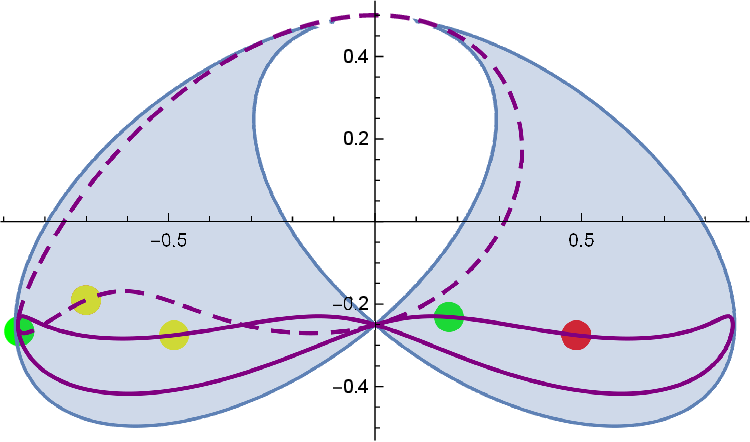}
\end{center}
\caption{Representation of surface $\widetilde{\cal T}_d$, in the bidimensional
space spanned by $(a^4_{\sm1},a^8_{\sm1})$, it is exactly the union of two ellipses
less their intersection. Four intersections of the ellipses, determined by
coefficients in (\ref{S1approx}), are found but they are not reliable.  We show
images of the same paths than in Fig.~\ref{surf0prime} but the second one is
trivial here.  Colored points are defined the same way as in Fig.~\ref{surf0},
here the conclusions written in  Fig.~\ref{surf0prime} only apply to the non
trivial path.}
\label{surf1}
\end{figure}

$\widetilde{\cal T}$ is embedded in a bidimensional space and reveals singular
points, which are the intersections of the two ellipses, shown in
Fig.~\ref{surf1}.  Confidence in the two-ellipse shape and in the properties of
$\widetilde{\cal T}$ (\ref{S1approx}) is complete, because our numerical
determination is actually exact. However, the coefficients in (\ref{S1approx})
must be improved because they lead to a wrong determination of the lower
intersection, which is exactly found from the $t'=0$ limit to be
$(0,{-}\frac14)$. Moreover, these coefficients give four intersections (among
which three are close to the upper boundary), where two seems more plausible.
Prescribing only two intersections would indeed settle new coefficients but we
haven't investigated this possibility because we could not prove any rigorous
founds.  Nevertheless, in case this prescription would give $(0,{-}\frac14)$ as
the lower limit, it is very plausible that it be correct.  

Apropos, one observes that $\widetilde{\cal T}_a$ (as well as
$\widetilde{\cal T}_d$) is a negative view of ${\cal R}'$. Since coefficients in
(\ref{S1approx}) lack of exactness, we are inclined to think that those in
(\ref{Rapprox}) too.

When mingling both surfaces, paths are characterized by winding number
$\omega_3$. On the contrary, encoding by 1 paths in $\widetilde{\cal T}_a$ and
by ${-}1$ those in $\widetilde{\cal T}_d$ exactly corresponds to
$-\omega_5$.  Altogether, $\widetilde{\cal T}$ is globally characterized
by $(\omega_3,-\omega_5)$. Eventually, one finds that the path is non trivial in
$\widetilde{\cal T}_a$ if $-\omega_3\omega_5>0$ and non trivial in 
$\widetilde{\cal T}_d$ if $-\omega_3\omega_5<0$.


\begin{thebibliography}{9}

\bibitem{Haldane} F. D. M. Haldane, Phys. Rev. Lett. \textbf{61}, 2015 (1988)

\bibitem{Bena} C. Bena \& G. Montambaux, New J. Phys. \textbf{11}, 095003
(2009)

\bibitem{FuchsEpjB} J.-N. Fuchs, F. Piéchon, M. O. Goerbig \& G. Montambaux,
Eur. Phys. J. B \textbf{77}, 351 (2010)

\bibitem{Lih2015} L.-K. Lim, J.-N. Fuchs \& G. Montambaux, Phys. Rev. A
\textbf{92}, 063627 (2015)

\bibitem{Read} N. Read \& D. Green, Phys. Rev. B \textbf{61}, 10267 (2000)

\bibitem{Kitaev} A. Y. Kitaev, Phys. Usp. \textbf{44}, 131 (2001)

\bibitem{FuKaneMele} L. Fu, C. L. Kane \& E. S. Mele, Phys. Rev. Lett.
\textbf{98}, 106803 (2007)

\bibitem{Dirac} P. A. R. Dirac, Proc. R. Soc. Lond. A \textbf{133}, 60 (1930)

\bibitem{AharonovBohm} Y. Aharonov \& D. Bohm, Phys. Rev. \textbf{115}, 485
(1959) 

\bibitem{Berry} M. V. Berry, Proc. R. Soc. Lond. A \textbf{392}, 45 (1984)

\bibitem{Volovik} G. E. Volovik, Sov. Phys. JETP \textbf{67}, 1804 (1988)

\bibitem{Bloore} F. J. Bloore, J. Phys. A: Math. Gen. \textbf{9}, 2059 (1976).

\bibitem{Poincare} J. Samuel \& R. Bhandari, Phys. Rev. Lett. \textbf{60}, 2339
(1988)

\bibitem{Arecchi} F. T. Arecchi, E. Courtens, R. Gilmore, H. Thomas, Phys. Rev.
A \textbf{6}, 2211 (1972)

\bibitem{Lieb} E. H. Lieb, Commun. Math. Phys. \textbf{31}, 327 (1973)

\bibitem{Montambaux} G. Montambaux, F. Piéchon, J.-N. Fuchs \& M. O. Goerbig,
Eur. Phys. J. B \textbf{72}, 509 (2009)

\bibitem{Lih2012} L.-K. Lim, J.-N. Fuchs \& G. Montambaux, Phys. Rev. Lett.
\textbf{108}, 175303 (2012)

\bibitem{Apaja} V. Apaja, M. Hyrkäs \& M. Manninen, Phys. Rev. A \textbf{82},
R041402 (2010)

\bibitem{Nathan} F. Nathan \& M. S. Rudner, New J. Phys. \textbf{17}, 125014
(2015)

\bibitem{Xiao} Y. Xiao, V. Pelletier, P. M. Chaikin \& D. A. Huse, Phys. Rev. B
\textbf{67}, 104505 (2003)

\bibitem{VolovB} G. E. Volovik, ``The Universe in a Helium Droplet'', 
(Oxford University Press, 2003)

\bibitem{Tsai} W.-F. Tsai, C. Fang, H. Yao \& J. Hu, New J. Phys. \textbf{17},
055016 (2015)

\bibitem{Lih-King} L.-K. Lim, J.-N. Fuchs, F. Piéchon \& G. Montambaux, Phys.
Rev. B \textbf{101}, 045131 (2020)

\bibitem{Goldman} N. Goldman, D.F. Urban \& D. Bercioux, Phys. Rev. A
\textbf{83}, 063601 (2011)

\bibitem{kagome} S. A. Owerre, J. Phys.: Condens. Matter \textbf{30}, 245803
(2018)

\bibitem{Chen} H. Chen, H. Nassar, G. L. Huang, J. Mech. Phys. Solids
\textbf{117}, 22 (2018)

\bibitem{Asano} K. Asano \& C. Hotta, Phys. Rev. B \textbf{83}, 245125
(2011)

\bibitem{ToulouseKleman} G. Toulouse \& M. Kléman, J. Physique Lett.
\textbf{37}, 149 (1976)

\bibitem{Goyal} S. K. Goyal, B. N. Simon, R. Singh \& S. Simon, J Phys. A: Math.
Theor. \textbf{49}, 165203 (2016)

\bibitem{Avron} J. E. Avron, R. Seiler \& B. Simon, Phys. Rev. Lett.
\textbf{51}, 51 (1983)

\bibitem{Lee} S.-Y. Lee, J.-H. Park, G. Go \& J. H. Han, J. Phys. Soc. Jpn.
\textbf{84}, 064005 (2015)

\bibitem{Allendoerfer} C. B. Allendoerfer, Amer. J. Math. \textbf{62}, 243
(1942)

\bibitem{Fenchel} W. Fenchel, J. London Math. Soc. \textbf{15}, 15 (1940)

\bibitem{Chern} S.-S. Chern, Ann. Math. \textbf{45}, 747 (1944)

\bibitem{prl100} P. Dietl, F. Piéchon \& G. Montambaux, Phys. Rev. Lett.
\textbf{100}, 236405 (2008)

\bibitem{Gamel} O. Gamel, Phys. Rev. A \textbf{93}, 062320 (2016)

\bibitem{Roth} L. M. Roth, Phys. Rev. \textbf{145}, 434 (1966)

\bibitem{Milnor} J. W. Milnor, ``\textit{Topology from the differentiable
viewpoint}'', (University Press of Virginia, 1965)

\bibitem{Parabolic} G. Montambaux, L.-K. Lim, J.-N. Fuchs \& F. Piéchon, Phys.
Rev. Lett. \textbf{121}, 256402 (2018)

\end{thebibliography}
\end{document}